\documentclass {IEEEtran}
\usepackage{bm}
\usepackage{epsfig}
\usepackage{stfloats}
\usepackage{graphicx}
\usepackage{subfigure}
\usepackage{epstopdf}
\usepackage{multirow}
\usepackage[sort, compress]{cite}
\usepackage{epsfig,graphics,subfigure}
\usepackage[cmex10]{amsmath}

\usepackage{array}
\usepackage{tabulary}
\usepackage{booktabs}
\usepackage{amsfonts}
\usepackage{algorithmic}
\usepackage{algorithm}

\usepackage{epstopdf}

\usepackage{amssymb}

\usepackage{textcomp}
\usepackage{xcolor}

\def\BibTeX{{\rm B\kern-.05em{\sc i\kern-.025em b}\kern-.08em
    T\kern-.1667em\lower.7ex\hbox{E}\kern-.125emX}}

\begin{document}
\bibliographystyle{IEEEtran}

\title{Movable Frequency Diverse Array for Wireless Communication Security}
\author{\IEEEauthorblockN{Zihao Cheng, Jiangbo Si, {\emph{Member, IEEE}}, Zan Li, {\emph{Senior Member, IEEE}}, \\
Pengpeng Liu, Yangchao Huang and Naofal Al-Dhahir, \emph{Fellow, IEEE}}}
 \maketitle
\begin{abstract}
Frequency diverse array (FDA) is a promising antenna technology to achieve physical layer security by varying the frequency of each antenna at the transmitter. However, when the channels of the legitimate user and eavesdropper are highly correlated, FDA is limited by the frequency constraint and cannot provide satisfactory security performance. In this paper, we propose a novel movable FDA (MFDA) antenna technology where the positions of antennas can be dynamically adjusted in a given finite region. Specifically, we aim to maximize the secrecy capacity by jointly optimizing the antenna beamforming vector, antenna frequency vector and antenna position vector. To solve this non-convex optimization problem with coupled variables, we develop a two-stage alternating optimization (AO) algorithm based on block successive upper-bound minimization (BSUM) method. Moreover, to evaluate the security performance provided by MFDA, we introduce two benchmark schemes, i.e., phased array (PA) and FDA. Simulation results demonstrate that MFDA can significantly enhance security performance compared to PA and FDA. In particular, when the frequency constraint is strict, MFDA can further increase the secrecy capacity by adjusting the positions of antennas instead of the frequencies.

\end{abstract}
\IEEEpeerreviewmaketitle

\section{Introduction}

Frequency diverse array (FDA) is a promising antenna technology used at the multi-antenna transmitter in wireless communization systems \cite{7087373,9440812,7817778,8672807,8752367}. Compared with conventional phased array (PA), FDA introduces a small frequency increment across array elements to yield distinct accumulating phase lags among array antennas and forms a special beampattern related to distance and angle. By tuning the antenna frequency vector (AFV) at the transmitter, FDA can enable the transmitted signals to constructively superimpose at the desired positions and destructively cancel at the undesired positions simultaneously \cite{7084678,6951408,1631800}. Especially in covert communication, FDA can achieve perfect covertness by optimizing AFV with the instantaneous channel state information (CSI) of the warden \cite{10097703,10008718,9905724}. Moreover, FDA develops an angle-range dependent beampattern which helps decouple the correlated channels to enhance physical layer security \cite{8078202,8104957,8458419}. Specifically, by optimizing AFV, the received signal levels of the legitimate user and eavesdropper can be distinguished clearly. When facing multi-eavesdropper scenario, a larger frequency increment is needed to decouple the correlated channels \cite{9133276,9560277}. However, the frequency increment is usually assumed to be much smaller than the carrier frequency due to receiver restriction which greatly limits FDA performance. For instance, when the legitimate and illegitimate users are proximal, such that their channels are highly correlated, the frequency increment needs to be greater than the carrier frequency to completely decouple their channels \cite{8078202,9905724}.

It is worth mentioning that both FDA and PA are fixed-distance arrays which means that the distance between each antenna is unchangeable and no smaller than a half-wavelength \cite{6798744,6736761,7416205}. Recently, movable antenna (MA) technology was proposed to achieve better communication performance where the antennas at the transmitter were movable within a confined region \cite{10243545,10286328,10318061,10278220,10304448,10354003,10464791,10382559}. By smartly designing the antenna positions vector (APV), MA can achieve higher spatial diversity gains by employing much fewer antennas compared with the conventional antenna selection (AS) technology \cite{10243545,10286328,10318061}. Specially, with known global instantaneous CSI, MA can achieve full array gain with null steering \cite{10278220}. Moreover, MA is capable of mitigating the interference from an adversary when MA is deployed at the receiver \cite{9369091}. In a multi-user and multi-interference scenario, MA can move in a three-dimensional (3D) region to obtain higher rate gain from the users and reduce the impact from different interference sources simultaneously. Besides, MA has great flexibility which can be deployed in one-dimensional (1D) line, two-dimensional (2D) surface and even six-dimensional (6D) region \cite{10304448,10354003,10464791,10382559}. In summary, MA outperforms fixed-distance antenna array and can be widely utilized in physical layer security.

In light of the above considerations, we propose a novel antenna technology, namely movable frequency diverse array (MFDA), which can further exploit the spatial degrees of freedom (DoFs) in transmit antennas based on FDA. Our main contributions are summarized as follows:

% In particular, when the channels of legitimate user and eavesdropper are highly correlated, FDA only optimizes AFV which is limited by the frequency offset constraint. By contrast, MFDA can optimize APV instead of optimizing AFV in this scenario to further improve the secrecy performance.

\begin{itemize}
\item[1.]

MFDA-aided multi-input single-output (MISO) wireless communication system is firstly investigated in this paper. Different from FDA, MFDA can not only adjust the frequency of each antenna but also adjust the position, such that it achieves two-dimensional security. Specifically, we jointly optimize the antenna beamforming vector (ABV), AFV and APV at the transmitter to maximize the secrecy capacity subject to the given constraints. Moreover, we investigate the influences of imperfect CSI at the eavesdropper on security performance.

\item[2.]

Since the formulated optimization problem is non-convex and has multiple coupled variables, we first reformulate it into a more tractable form. Then, we propose a low-complexity two-stage alternating optimization (AO) algorithm to solve this problem. Specifically, in the first stage, we alternately optimize AFV and APV by utilizing the block successive upper-bound minimization (BSUM) method. In the second stage, with optimized AFV and APV, the optimal ABV expression can be obtained in closed form.

\item[3.]

In the case of imperfect CSI, we assume that the exact coordinate of the eavesdropper is unknown, but the suspicious region of the eavesdropper is known. We first divide the region into multiple discrete samples and consider the worst sample. Then, the proposed two-stage AO algorithm is utilized to solve the problem. Different from the case of perfect CSI, the block coordinate descent (BCD) method replaces the BSUM to optimize AFV and APV in the first stage, and the AO and CVX method is utilized to optimize ABV in the second stage.

\item[4.]

MFDA can achieve better security performance than FDA and PA. Specifically, when the channels of the legitimate user and eavesdropper are highly correlated, FDA is limited by the frequency constraint and cannot provide satisfactory performance. By contrast, MFDA can adjust the positions of antennas instead of the frequencies to further improve the performance.

\end{itemize}

The rest of this paper is organized as follows. Section II describes the MFDA-aided MISO system model and formulates the optimization problem.  Section III develops a two-stage AO algorithm to solve the problem and investigates the influences of beamformer refresh frequency on secrecy capacity. Section VI considers the case of imperfect CSI at the eavesdropper. Numerical results are presented and discussed in Section V, and conclusions are drawn in Section IV.

\emph{Notations}: Boldface uppercase and lowercase letters denote matrices and vectors, respectively. Italic letters denote scalars; ${\bf{x}}^\dag $ denotes the Hermitian of a vector  $\bf{x} $. For a given matrix $\bf{A}$, we denote its transpose, Hermitian and inverse by ${\bf{A}}^T$, ${\bf{A}}^{\dag}$ and ${\bf{A}}^{-1}$, respectively. Denote ${\left\| . \right\|^2}$ as ${\ell _2}$-norm. $\mathbb{C}^{M \times N}$ denotes the complex $M \times N$ matrices. $x \sim {\cal{CN}}\left( {\bf{\Phi} ,\bf{\Omega}} \right)$ denotes the circular symmetric complex Gaussian random vector with mean vector $\bf{\Phi}$ and covariance matrix $\bf{\Omega}$. $ \left\langle { \cdot , \cdot } \right\rangle $ denotes the inner product. ${\bf{tr}} \left(  \cdot  \right)$ is trace operator of a matrix. $\left\lfloor  \cdot  \right\rfloor $ and $\left\lceil  \cdot  \right\rceil $ are the nearest integer towards $ - \infty $ and $ + \infty $, respectively. $c = a{\kern 1pt} \bmod {\kern 1pt} b$ means $c$ is the remainder of integer $a$ divided by integer $b$.

\section{System Model}

\begin{figure}[htbp]
\centering
\subfigure[]{
\includegraphics[width=1.5in]{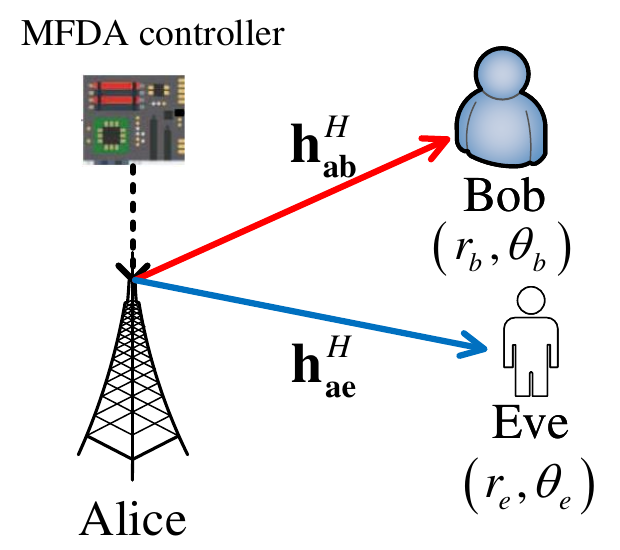}
\label{Fig1-1}}
\quad
\subfigure[]{
\includegraphics[width=1.5in]{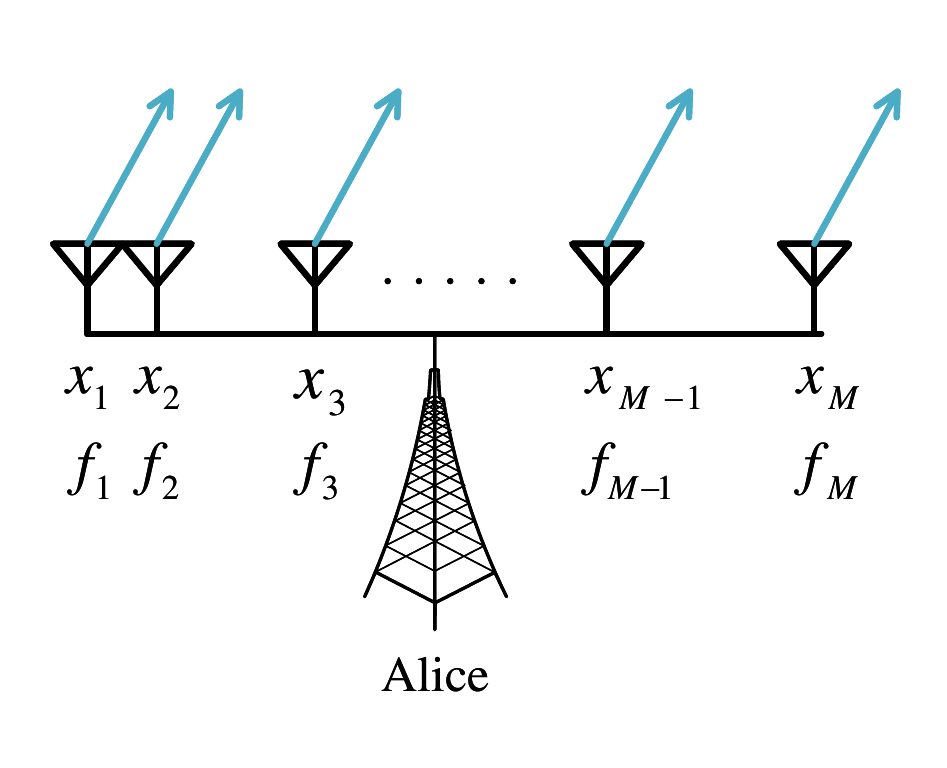}
\label{Fig1-2}}
\caption{MFDA-aided MISO wireless communication system model}\label{Fig1}
\end{figure}

\subsection{System Description and Channel Model}

In this paper, we consider a MISO wireless communication system where 1D linear MFDA is deployed to assist in the communication from the legitimate user Alice to Bob, while preventing being eavesdropped by the illegitimate user Eve. The number of antennas at Alice is denoted by $M$, and both Bob and Eve are assumed to be equipped with a single antenna. As shown in Fig. 1(a), a smart MFDA controller helps adjust APV ${\bf{x}}\buildrel \Delta \over = {\left[ {{x_1},...,{x_M}} \right]} \in {{\mathbb{C}}^{1 \times M}}$ and AFV ${\bf{f}} \buildrel \Delta \over = {\left[ {{f_1},...,{f_M}} \right]} \in {{\mathbb{C}}^{1 \times M}}$ at Alice via a separate wireless link. Compared with PA, MFDA introduces small frequency increments across the array antennas, such that the frequency of the $m$-th antenna is given by
\begin{align}\label{radiation_frequency}\
{f_m} = {f_C} + \Delta {f_m},m = 1,...,M,
\end{align}
where $f_C$ is the carrier frequency and $\Delta {f_m}$ is the introduced frequency increment. We assume that $0 \le \Delta {f_m} \le \Delta F$ and the frequency increment is much smaller than the carrier frequency $\Delta F \ll {f_C}$.

Without loss of generality, we define the first antenna of Alice as the origin of the coordinate. Thus, the coordinates of Bob and Eve are denoted by $\left( {{r_b},{\theta _b}} \right)$ and $\left( {{r_e},{\theta _e}} \right)$, respectively. To characterize the theoretical performance gain brought by the MFDA, we assume that $\left( {{r_b},{\theta _b}} \right)$ and $\left( {{r_e},{\theta _e}} \right)$ are perfectly known by Alice and all channels are line-of-sight (LoS) channels as in \cite{8078202,8104957,8458419}. In addition, the channels of Alice$\to$Bob and Alice$\to$Eve are defined as
\begin{align}\label{h_ab}\
{{\bf{h}}_{ab}}\left( {t,{r_b},{\theta _b}} \right) = {\rm{Lfs}}\left( {{r_b}} \right){\left[ {{h_1}\left( {t,{r_b},{\theta _b}} \right),...,{h_M}\left( {t,{r_b},{\theta _b}} \right)} \right]^T}
\end{align}
and
\begin{align}\label{h_ae}\
{{\bf{h}}_{ae}}\left( {t,{r_e},{\theta _e}} \right) = {\rm{Lfs}}\left( {{r_e}} \right){\left[ {{h_1}\left( {t,{r_e},{\theta _e}} \right),...,{h_M}\left( {t,{r_e},{\theta _e}} \right)} \right]^T},
\end{align}
where ${h_m}\left( {t,{r_b},{\theta _b}} \right) = {e^{ - j2\pi {f_m}\left[ {t - \frac{{{r_b} - {x_m}\sin {\theta _b}}}{c}} \right]}}$ and ${h_m}\left( {t,{r_e},{\theta _e}} \right)= {e^{ - j2\pi {f_m}\left[ {t - \frac{{{r_e} - {x_m}\sin {\theta _e}}}{c}} \right]}}$, $\forall m$. $c$ is the speed of light and ${\rm{Lfs}}\left( \cdot  \right)$ is the path loss factor.

\subsection{Problem Formulation}

We observe from \eqref{h_ab} and \eqref{h_ae} that the channels of MFDA are dependent on time $t$. Thus, the ABV at Alice is assumed to be time-varying to adapt to the variable channels which can be defined as
\begin{align}\label{beamforming}\
{\bf{w}}\left( t \right) = {\left[ {{w_1}\left( t \right),...,{w_M}\left( t \right)} \right]^T}.
\end{align}
Then, the received signals at Bob and Eve can be written as
 \begin{align}\label{y_b}\
{{\bf{y}}_b}\left( t \right) = {\bf{h}}_{ab}^H\left( {t,{r_b},{\theta _b}} \right){\bf{w}}\left( t \right) + {{\bf{n}}_b}
\end{align}
and
 \begin{align}\label{y_b}\
{{\bf{y}}_e}\left( t \right) = {\bf{h}}_{ae}^H\left( {t,{r_e},{\theta _e}} \right){\bf{w}}\left( t \right) + {{\bf{n}}_e},
\end{align}
where ${{\bf{n}}_b}$ and ${{\bf{n}}_e}$ are the additive white Gaussian noise (AWGN) at Bob and Eve with mean zero and variances $\sigma _b^2$ and $\sigma _e^2$, i.e., ${{\bf{n}}_b} \sim {\cal{CN}}\left( {0,\sigma _b^2} \right)$ and ${{\bf{n}}_e} \sim {\cal{CN}}\left( {0,\sigma _e^2} \right)$. Thus, the instantaneous achievable rates at Bob and Eve are given by
 \begin{align}\label{R_b}\
{R_b}\left( t \right) = \log \left( {1 + \sigma _b^{ - 2}{{\left| {{\bf{h}}_{ab}^H\left( {t + \frac{{{r_b}}}{c},{r_b},{\theta _b}} \right){\bf{w}}\left( t \right)} \right|}^2}} \right)
\end{align}
and
 \begin{align}\label{R_e}\
{R_e}\left( t \right) = \log \left( {1 + \sigma _e^{ - 2}{{\left| {{\bf{h}}_{ae}^H\left( {t + \frac{{{r_e}}}{c},{r_e},{\theta _e}} \right){\bf{w}}\left( t \right)} \right|}^2}} \right).
\end{align}
Next, the instantaneous secrecy capacity can be expressed as
 \begin{align}\label{C_sec}\
{C_{\sec }}\left( t \right) &= {R_b}\left( t \right) - {R_e}\left( t \right)\nonumber\\
 &= \log \left( {\frac{{1 + \sigma _b^{ - 2}{{\left| {{\bf{h}}_{ab}^H\left( {t + \frac{{{r_b}}}{c},{r_b},{\theta _b}} \right){\bf{w}}\left( t \right)} \right|}^2}}}{{1 + \sigma _e^{ - 2}{{\left| {{\bf{h}}_{ae}^H\left( {t + \frac{{{r_e}}}{c},{r_e},{\theta _e}} \right){\bf{w}}\left( t \right)} \right|}^2}}}} \right).
\end{align}

To improve the secrecy performance gain, we aim to maximize the instantaneous secrecy capacity by jointly optimizing ABV, APV and AFV at Alice with the perfect global CSI. The optimization problem can be formulated as follows:
\begin{subequations}
\begin{align}\label{OP}\
&\mathop {\max }\limits_{\left\{ {{\bf{w}}\left( t \right), {\bf{x}}, {\bf{f}}  } \right\}} {\rm{ }}{C_{\sec }}\left( t \right)\\
&{s.t.} \quad {\left\| {\bf{w}}\left( t \right) \right\|^2} \le {P_{max}},\label{power_constraintop}\\
&\quad \quad {x_1} = 0,{x_M} \le {D_{\max }}, \label{distance_constraint1op}\\
&\quad \quad {x_m} - {x_{m - 1}} \ge {D_0}, \forall m, \label{distance_constraint2op}\\
&\quad \quad {f_C} \le {f_m} \le {f_C} + \Delta F, \forall m , \label{frequency_constraintop}
\end{align}
\end{subequations}
where \eqref{power_constraintop} is the maximum power constraint and ${P_{max}}$ is the maximum transmit power. \eqref{distance_constraint1op} is 1D linear range constraint which ensures that MFDA is moved within the predetermined region $\left[ {0,{D_{\max }}} \right]$ and ${{D_{\max }}}$ is the movable boundary of MFDA. \eqref{distance_constraint2op} aims to avoid antenna coupling and ${D_0}$ is the minimum distance between any adjacent antennas of MFDA. Besides, \eqref{frequency_constraintop} is the frequency constraint of MFDA.

\section{Proposed Algorithm for Optimization Problem}

We observe that the optimization problem \eqref{OP} contains three optimized variables which are
 are tightly coupled and make the problem challenging to solve. Usually, AO algorithm is utilize to solve this problem \cite{8972400,9151964}. However, when there are multiple variables, the complexity of AO algorithm greatly increases and the convergence accuracy decreases. To avoid these problems, we propose a low-complexity two-stage AO algorithm to solve problem (10).

\subsection{Optimizing ${\bf{w}}\left( t \right)$ with Given ${\bf{x}}$ and ${\bf{f}}$ }

According to \eqref{C_sec}, problem (10) can be reformulated as
\begin{subequations}
\begin{align}\label{OP1}\
&\mathop {\max }\limits_{{\bf{w}}\left( t \right)} {\rm{ }}\frac{{1 + {{\bf{w}}^H}\left( t \right){\bf{A}} {\bf{w}}\left( t \right)}}{{1 + {{\bf{w}}^H}\left( t \right){\bf{B}}{\bf{w}}\left( t \right)}}\\
&{s.t.} \quad {{\bf{w}}^H}\left( t \right){\bf{w}}\left( t \right) \le {P_{\max }},\label{power_constraint}
\end{align}
\end{subequations}
where ${\bf{A}} = \frac{1}{{\sigma _b^2}}{{\bf{h}}_{ab}}\left( {t + \frac{{{r_b}}}{c},{r_b},{\theta _b}} \right){\bf{h}}_{ab}^H\left( {t + \frac{{{r_b}}}{c},{r_b},{\theta _b}} \right)$ and ${\bf{B}}  = \frac{1}{{\sigma _e^2}}{{\bf{h}}_{ae}}\left( {t + \frac{{{r_e}}}{c},{r_e},{\theta _e}} \right){\bf{h}}_{ae}^H\left( {t + \frac{{{r_e}}}{c},{r_e},{\theta _e}} \right)$. The optimal ABV is given by \cite{5485016}
 \begin{align}\label{w_opt}\
{{\bf{w}}_{opt}}\left( t \right) = \sqrt {{P_{\max }}} {{\bf{u}}_{\max }},
\end{align}
where ${{\bf{u}}_{\max }}$ is the normalized eigenvector corresponding to the largest eigenvalue of the matrix ${\left( {{\bf{B}} + P_{\max }^{ - 1}{{\bf{I}}_M}} \right)^{ - 1}}\left( {{\bf{A}}  + P_{\max }^{ - 1}{{\bf{I}}_M}} \right)$.

By substituting ${{\bf{w}}_{opt}}\left( t \right)$ into \eqref{C_sec}, we rewrite problem (10) into a more tractable form as
\begin{subequations}
\begin{align}\label{OP2}\
&\mathop {\min }\limits_{\left\{ {{\bf{x}},{\bf{f}}} \right\}} {\left| {\left\langle {{{\bf{h}}_{ab}}\left( {t,{r_b},{\theta _b}} \right),{{\bf{h}}_{ae}}\left( {t,{r_e},{\theta _e}} \right)} \right\rangle } \right|^2}\\
&{s.t.} \quad {x_1} = 0,{x_M} \le {D_{\max }}, \label{distance_constraint1}\\
&\quad \quad {x_m} - {x_{m - 1}} \ge {D_0}, \forall m, \label{distance_constraint2}\\
&\quad \quad {f_C} \le {f_m} \le {f_C} + \Delta F, \forall m . \label{frequency_constraint}
\end{align}
\end{subequations}
The detailed reformulation proof is presented in Appendix A. According to \eqref{h_ab} and \eqref{h_ae}, we have
 \begin{align}\label{habwithhae}\
&{\left| {\left\langle {{{\bf{h}}_{ab}}\left( {t,{r_b},{\theta _b}} \right),{{\bf{h}}_{ae}}\left( {t,{r_e},{\theta _e}} \right)} \right\rangle } \right|^2}\nonumber\\
 &= {\left| {\sum\limits_{m = 1}^M {{e^{j2\pi {f_m}\left( {{x_m}\frac{{\sin {\theta _e} - \sin {\theta _b}}}{c} + \frac{{{r_b} - {r_e}}}{c}} \right)}}} } \right|^2}\nonumber\\
 &= \sum\limits_{m = 1}^M {\sum\limits_{n = 1}^M {{e^{j2\pi \left( {{f_m}{\tau _m} - {f_n}{\tau _n}} \right)}}} }  \nonumber\\
 &= M + {\sum\limits_{m = 1}^{M} {{\sum\limits_{\scriptstyle{n=1}\hfill\atop
\scriptstyle{n \ne  m}\hfill}^{M} {{\cos \left[ {2\pi \left( {{f_m}{\tau _m} - {f_n}{\tau _n}} \right)} \right]}}  }} }.
\end{align}
where ${\tau _m} = {x_m}\frac{{\sin {\theta _e} - \sin {\theta _b}}}{c} + \frac{{{r_b} - {r_e}}}{c}$. Next, problem (13) can be reformulated as
\begin{subequations}
\begin{align}\label{OP3}\
&\mathop {\min }\limits_{\left\{ {{\bf{x}},{\bf{f}}} \right\}} {\sum\limits_{m = 1}^{M} {{\sum\limits_{\scriptstyle{n=1}\hfill\atop
\scriptstyle{n \ne  m}\hfill}^{M} {\cos \left[ {2\pi \left( {{f_m}{\tau _m} - {f_n}{\tau _n}} \right)} \right]} }} }\\
&{s.t.} \quad {x_1} = 0,{x_M} \le {D_{\max }}, \label{distance_constraint1}\\
&\quad \quad {x_m} - {x_{m - 1}} \ge {D_0}, \forall m, \label{distance_constraint2}\\
&\quad \quad {f_C} \le {f_m} \le {f_C} + \Delta F, \forall m. \label{frequency_constraint111}
\end{align}
\end{subequations}

{\emph{Remark 1:}}
After utilizing the optimal ABV in \eqref{w_opt}, \eqref{OP2} reduces to minimizing the inner product between ${{\bf{h}}_{ab}}\left( {t,{r_b},{\theta _b}} \right)$ and ${{\bf{h}}_{ae}}\left( {t,{r_e},{\theta _e}} \right)$. This implies that the adjusted APV and AFV aim to make the channels of Bob and Eve as orthogonal as possible. In particular, the ideal case is ${{\bf{h}}_{ab}}\left( {t,{r_b},{\theta _b}} \right) \perp {{\bf{h}}_{ae}}\left( {t,{r_e},{\theta _e}} \right)$. Unfortunately, due to the frequency constraint in \eqref{frequency_constraint111}, the ideal case is difficult to achieve \cite{8078202}. By contrast, MFDA introduces a new dimension. When the frequency constraint is strict, MFDA can optimize APV to further improve the security performance instead of optimizing AFV.

{\emph{Remark 2:}}
According to problem (15), it is clear that ABV cannot affect the optimization of APV or AFV. It means that ABV is not coupled with APV and AFV, but APV and AFV are coupled with each other. Thus, problem (10) can be recast to a two-stage algorithm. Specifically, the first stage involves obtaining the optimal APV and AFV in problem (15) using the AO algorithm. The second stage is obtaining the optimal ABV by \eqref{w_opt}. Based on this fact, we propose a two-stage AO algorithm in this paper. Moreover, it is worth mentioning that the optimal APV and AFV are independent on time, but the optimal ABV is dependent on time.

\subsection{Optimizing ${\bf{x}}$ with Given ${\bf{w}}\left( t \right)$ and ${\bf{f}}$ }

With given ${\bf{f}}$, problem (15) can be rewritten as
\begin{subequations}
\begin{align}\label{OP4}\
&\mathop {\min }\limits_{\bf{\tau }} {\sum\limits_{m = 1}^{M} {{\sum\limits_{\scriptstyle{n=1}\hfill\atop
\scriptstyle{n \ne  m}\hfill}^{M} \cos \left[ {2\pi \left( {{\tau _m}{f_m} - {\tau _n}{f_n}} \right)} \right]  }} }\\
&{s.t.} \quad {x_1} = 0,{x_M} \le {D_{\max }}, \label{distance_constraint1}\\
&\quad \quad {x_m} - {x_{m - 1}} \ge {D_0}, \forall m, \label{distance_constraint2}
\end{align}
\end{subequations}
where ${\tau}  = \left\{ {{\tau _{1,}}{\tau _{2,...,}}{\tau _M}} \right\}$. Problem \eqref{OP4} is clearly non-convex, thus we resort to the BSUM method to solve it. BSUM solves the problem by formulating a new convex function to approach the optimization target. More specifically, in the $s$-th iteration of BSUM, we first optimize ${{{\tau _m}}}$ where $m = s{\kern 1pt} {\kern 1pt} {\kern 1pt} {\kern 1pt} {\rm{mod}}{\kern 1pt} {\kern 1pt} {\kern 1pt} M$, and obtain the optimal position $\tau_m^{opt}$. Then, we update ${{\bf{\tau}}^{s - 1}}$ with $\tau_m^{opt}$, i.e., ${{\bf{\tau}}^s} = {{\bf{\tau}}^{s - 1}}$. The problem can be written as
\begin{subequations}
\begin{align}\label{OPAO1}\
&\mathop {\min }\limits_{{\tau_m}} \quad { {\sum\limits_{\scriptstyle{n=0}\hfill\atop
\scriptstyle{n \ne  m}\hfill}^{M-1} {{\cos \left[ {2\pi \left( {{\tau_m}{f _m} - {\tau_n^{s - 1}}{f_n}} \right)} \right]}} }}\\
&{s.t.} \quad {x_1} = 0,{x_M} \le {D_{\max }}, \label{distance_constraint1}\\
&\quad \quad {x_m} - {x_{m - 1}} \ge {D_0}, \forall m. \label{distance_constraint2}
\end{align}
\end{subequations}
Next, we define the non-convex function in \eqref{OPAO1} as
\begin{align}\label{oldobject}\
{y_{m,n}}\left( {{x}} \right) = \cos \left[ {2\pi \left( {{x}{f _m} - {\tau_n^{s - 1}}{f _n}} \right)} \right].
\end{align}
Following the BSUM method, the objective function \eqref{oldobject} is approximated by the upper-bound quadratic function ${u_{m,n}}\left( {{x}} \right)$, which is defined by
\begin{align}\label{newobject}\
{u_{m,n}}\left( {{x}} \right) = {k_{m,n}}{\left( {{x} - {\zeta _{m,n}}} \right)^2} + {\delta _{m,n}},
\end{align}
where ${k_{m,n}} \in \mathbb{R}$, ${k_{m,n}} >0 $, ${{\zeta _{m,n}}}\in \mathbb{R}$ and ${\delta _{m,n}}\in \mathbb{R}$ are the parameters of the new quadratic function. For a given point ${\tau_m^{s - 1}}$ in \eqref{oldobject}, the approximate function \eqref{newobject} should satisfy the following constraints:
\begin{align}\label{newobject_constraints}\
\left\{ {\begin{array}{*{20}{l}}
{{u_{m,n}}\left( {\tau_m^{s - 1};\tau_n^{s - 1}} \right) = {y_{m,n}}\left( {\tau_m^{s - 1};\tau_n^{s - 1}} \right)}\\
{u_{m,n}^{'}\left( {\tau_m^{s - 1};\tau_n^{s - 1}} \right) = y_{m,n}^{'}\left( {\tau_m^{s - 1};x_n^{s - 1}} \right)}\\
{{u_{m,n}}\left( {{\zeta _{m,n}};\tau_n^{s - 1}} \right) \in \left\{ {1, - 1} \right\}},
\end{array}} \right.
\end{align}
Next, we can obtain the parameters of the new quadratic function in \eqref{newobject} by utilizing Lemma 1.

{\emph{\textbf{Lemma 1:}}}
If $y_{m,n}^{'} \left( {\tau_m^{s - 1};\tau_n^{s - 1}} \right) \ne 0$, we have
\begin{align}\label{nle0kxidelta}\
\left\{ {\begin{array}{*{20}{l}}
{{k_{m,n}} = \frac{{ - \pi {f_m}\sin \left[ {2\pi \left( {\tau_m^{s - 1}{f_m} - \tau_n^{s - 1}{f_n}} \right)} \right]}}{{\tau_m^{s - 1} - {\zeta _{m,n}}}}}\\
{{\zeta _{m,n}} = \left\{ {\begin{array}{*{20}{l}}
{\frac{{\left\lfloor {2\left( {\tau_m^{s - 1}{f_m} - \tau_n^{s - 1}{f_n}} \right)} \right\rfloor }}{{2{f_m}}} + \frac{{\tau_n^{s - 1}{f_n}}}{{{f_m}}}}\\
{if \quad y_{m,n}^{'}\left( {\tau_m^{s - 1};\tau_n^{s - 1}} \right) > 0}\\
{\frac{{\left\lceil {2\left( {\tau_m^{s - 1}{f_m} - \tau_n^{s - 1}{f_n}} \right)} \right\rceil }}{{2{f_m}}} + \frac{{\tau_n^{s - 1}{f_n}}}{{{f_m}}}}\\
{if \quad y_{m,n}^{'}\left( {\tau_m^{s - 1};\tau_n^{s - 1}} \right) < 0}
\end{array}} \right.}\\
{\delta _{m,n}} = \cos \left[ {2\pi \left( {\tau_m^{s - 1}{f_m} - \tau_n^{s - 1}{f_n}} \right)} \right]\\
 \quad  \quad  - {k_{m,n}}{\left( {\tau_m^{s - 1} - {\zeta _{m,n}}} \right)^2}.
\end{array}} \right.
\end{align}
If $y_{m,n}^{'}\left( {\tau_m^{s - 1};\tau_n^{s - 1}} \right) = 0$, we have
\begin{align}\label{n0kxidelta}\
\left\{ {\begin{array}{*{20}{l}}
{{k_{m,n}} =  - 2{\pi ^2}f_m^2{y_{m,n}}\left( {\tau_m^{s - 1};\tau_n^{s - 1}} \right)}\\
{{\zeta _{m,n}} = \tau_m^{s - 1}}\\
{{\delta _{m,n}} = {y_{m,n}}\left( {\tau_m^{s - 1};\tau_n^{s - 1}} \right) \in \left\{ { - 1,1} \right\}}.
\end{array}} \right.
\end{align}
The proof of Lemma 1 is presented in Appendix B. Therefore problem \eqref{OPAO1} can be replaced by
\begin{subequations}
\begin{align}\label{OPAO1_122}\
&\mathop {\min }\limits_{{\tau_m}} \quad { {\sum\limits_{\scriptstyle{n=0}\hfill\atop
\scriptstyle{n \ne  m}\hfill}^{M-1} {k_{m,n}}{\left( {\tau_m - {\zeta _{m,n}}} \right)^2} + {\delta _{m,n}} }}
\end{align}
\end{subequations}
It is clear that the optimal solution of problem \eqref{OPAO1_122} is given by
\begin{align}\label{optx}\
\tau _m^{opt} = \frac{ { {\sum\limits_{\scriptstyle{n=0}\hfill\atop
\scriptstyle{n \ne  m}\hfill}^{M-1} {{{k_{m,n}}{\zeta _{m,n}}}} }}}{ { {\sum\limits_{\scriptstyle{n=0}\hfill\atop
\scriptstyle{n \ne  m}\hfill}^{M-1} {{{k_{m,n}}}} }}}.
\end{align}
Thus, the optimal position of the $m$-th antenna is given by
\begin{align}\label{optxtau}\
x_m^{opt} = \left( {\tau _m^{opt} - \frac{{{r_b} - {r_e}}}{c}} \right){\left( {\sin {\theta _e} - \sin {\theta _b}} \right)^{ - 1}}c.
\end{align}
According to the constraints \eqref{distance_constraint1} and \eqref{distance_constraint2}, the optimal position of the $m$-th antenna is given by
 \begin{align}\label{opt_xm}\
x_m^{opt} = \left\{ {\begin{array}{*{20}{l}}
0&{m = 1}\\
{x_{m - 1}^s + {D_0}}&{{x^{opt}} \le x_{m - 1}^s + {D_0}}\\
{{x^{opt}}}&{{D_1} \ge {x^{opt}} > x_{m - 1}^s + {D_0}}\\
{{D_1}}&{{D_1}< {x^{opt}}.}
\end{array}} \right.
\end{align}
where ${D_1} = {D_{\max }} - \left( {M - m} \right){D_0}$. The convergence of the BSUM is verified in Fig. 2.

\subsection{Optimizing ${\bf{f}}$ with Given ${\bf{w}}\left( t \right)$ and ${\bf{x}}$ }

With given ${\bf{w}}\left( t \right)$ and ${\bf{x}}$, the optimization problem can be formulated as
\begin{subequations}
\begin{align}\label{OPAO2}\
&\mathop {\min }\limits_{\bf{f}} {\sum\limits_{m = 1}^{M} {{\sum\limits_{\scriptstyle{n=1}\hfill\atop
\scriptstyle{n \ne  m}\hfill}^{M} \cos \left[ {2\pi \left( {{\tau_m}{f_m} - {\tau_n}{f_n}} \right)} \right]  }} }\\
&{s.t.} \quad {f_C} \le {f_m} \le {f_C} + \Delta F, \forall m . \label{frequency_constraint}
\end{align}
\end{subequations}
Problem (26) is non-convex which can be solved by the BSUM method developed in the previous subsection. Due to the similar mathematical derivation, we omit the proof here for brevity. The optimal frequency of the $m$-th antenna is given by
 \begin{align}\label{opt_fm}\
f_m^{opt} = \left\{ {\begin{array}{*{20}{l}}
{{f_C}}&{{f_{opt}} < {f_C}}\\
{{f_{opt}}}&{{f_C} \le {f_{opt}} < {f_C} + \Delta F}\\
{{f_C} + \Delta F}&{{f_C} + \Delta F \le {f_{opt}}.}
\end{array}} \right.
\end{align}
where ${{f_{opt}}}$ can be obtained by Lemma 1.

\subsection{Overall Two-Stage AO Algorithm}

Inspired by {\emph{Remark 1}}, we propose a two-stage AO algorithm with low complexity. Specifically, the first stage aims to alternately optimize the APV and AFV by utilizing the BSUM method which is summarized in Algorithm 1. The second stage is to obtain the optimal ABV with the optimized APV and AFV. The overall algorithm is summarized in Algorithm 2.

\begin{algorithm}
\caption{Optimizing ${\bf{x}}$ by BSUM}
\begin{algorithmic}[1]
\STATE {\bf{Initialization:}} Initializing ${\bf{f}}$ via Algorithm 1 .
\STATE {\bf{Set:}} $s=0$.
\STATE {{\bf{Repeat:}}\\
        a) $s=s+1$, and $m = s{\kern 1pt} {\kern 1pt} {\kern 1pt} {\kern 1pt} {\rm{mod}}{\kern 1pt} {\kern 1pt} {\kern 1pt} {\rm{M}}$.\\
        b) The new quadratic function ${u_{m,n}}\left( {{x}} \right)$ can be obtained by Lemma 1. Then, the optimal position of $m$-th antenna can be obtained by \eqref{opt_xm}.\\
        c) We update ${{\bf{x}}^{s-1}}$ with $x_m^{opt}$ and obtain ${{\bf{x}}^s}$.\\}
\STATE {{\bf{Until}}  $\left| {{C_{\sec }}\left( {{{\bf{w}}_{opt}},{{\bf{x}}^s},{\bf{f}}} \right) - {C_{\sec }}\left( {{{\bf{w}}_{opt}},{{\bf{x}}^{s - 1}},{\bf{f}}} \right)} \right| \le \varsigma $ is satisfied where $\varsigma$ is a small convergence value.}
\end{algorithmic}
\end{algorithm}

\begin{algorithm}
\caption{Overall two-stage AO algorithm}
\begin{algorithmic}[1]
\STATE {\bf{Initialization:}} Initializing ${{{\bf{w}}_{opt}}}$ via \eqref{w_opt}.
\STATE {\bf{Set:}} $i=0$.
\STATE {{\bf{Repeat:}}\\
        a) $i=i+1$.\\
        b) Optimizing  ${\bf{x}}$ and ${\bf{f}}$ by BSUM.\\
        c) Obtaining  ${{{\bf{w}}_{opt}}}$ with the optimized ${{\bf{x}}^s}$ and ${{\bf{f}}^s}$.\\}
\STATE {{\bf{Until}} $\left| {{C_{\sec }}\left( {{{\bf{w}}_{opt}},{{\bf{x}}^s},{{\bf{f}}^s}} \right) - {C_{\sec }}\left( {{{\bf{w}}_{opt}},{{\bf{x}}^{s - 1}},{{\bf{f}}^{s - 1}}} \right)} \right| \le \varsigma $ is satisfied.}
\end{algorithmic}
\end{algorithm}

The complexity of Algorithm 1 is ${\cal{O}}\left( {{I_{BSUM}}} \right)$ where ${{I_{BSUM}}}$ denotes the number of iterations required for BSUM convergence. The overall complexity of Algorithm 2 is ${\cal{O}}\left( {{I_{AO}}\left( {{I_{BSUM}} + {I_{BSUM}}} \right)} \right)$ where ${{I_{AO}}}$ denotes the number of iterations required for AO algorithm convergence.

\subsection{Beamformer Refresh Frequency}

According to \eqref{OP}, the formulated optimization problem is an instantaneous problem and the optimal solution is dependent on time. Besides, from {\emph{Remark 2}}, we know that the optimized AFV and APV are independent on time, and only the optimized ABV is dependent on time. Based on these facts, the optimal solution is related to the speed of ABV optimization or the so-called beamformer refresh frequency (BRF) \cite{8078202}. In particular, if BRF can keep up with channel refresh frequency (CRF), the instantaneous secrecy capacity can be maximized. However, due to the device restriction and implementation cost, high beamformer refresh is difficult to achieve. Hence, it is necessary to investigate the influences of BRF and CRF on security performance.

More specifically, we define BRF and CRF as $\frac{1}{{\Delta {T_{BRF}}}}$ and $\frac{1}{{\Delta {T_{CRF}}}}$, respectively, where $T$ is defined as the whole transmission time. Then, the channel refresh time points are given by $t_l^c = \left( {l - 1} \right)\frac{1}{L},l = 1,2,...,L$ where $L = \frac{T}{{\Delta {T_{CRF}}}}$, and the beamformer refresh time points are given by $t_k^w = \left( {k - 1} \right)\frac{1}{K},k = 1,2,...,K$ where $K = \frac{T}{{\Delta {T_{BRF}}}}$. Thus, the secrecy capacity is given by
 \begin{align}\label{sc}\
{C_{\sec }} &= \frac{1}{T}\int_0^{T - 1} {{C_{\sec }}\left( {{{\bf{w}}_{opt}}\left( t \right),{{\bf{x}}_{opt}},{{\bf{f}}_{opt}}} \right)dt} \nonumber\\
 &= \frac{1}{{LK}}\sum\limits_{l = 1}^L {\sum\limits_{k = 1}^K {{{\log }_2}} } \left( {\frac{{1 + {{\bf{w}}^H}\left( {t_k^w} \right){\bf{A}}\left( {t_l^c} \right){\bf{w}}\left( {t_k^w} \right)}}{{1 + {{\bf{w}}^H}\left( {t_k^w} \right){\bf{B}}\left( {t_l^c} \right){\bf{w}}\left( {t_k^w} \right)}}} \right).
\end{align}

\section{Imperfect CSI at Eve}

In this section, we consider a more practical scenario where Alice has imperfect CSI knowledge about Eve. According to \eqref{h_ae}, we conclude that the imperfect CSI at Eve is mainly caused by an estimation error in Eve's range and angle. Without loss of generality, we assume that the exact coordinate of Eve is unknown, but the suspicious region for Eve is known. The uncertainty regions in Eve's range and angle is denoted as ${\cal{R}} \buildrel \Delta \over = \left[ {{r_e} - \Delta r,{r_e} + \Delta r} \right]$ and ${\cal{A}} \buildrel \Delta \over = \left[ {{\theta _e} - \Delta \theta ,{\theta _e} + \Delta \theta } \right]$, where ${\Delta r}$ and ${\Delta \theta }$ is the estimation error of the range and angle, respectively. Since ${\cal{R}}$ and ${\cal{A}}$ are continuous sets, we propose to uniformly generate some discrete samples over the sets which are given as ${r_y} = {r_e} - \Delta r + \left( {y - 1} \right)\frac{{2\Delta r}}{{Y - 1}},y \in \left\{ {1,2,...,Y} \right\}$ and ${\theta _z} = {\theta _e} - \Delta \theta  + \left( {z - 1} \right)\frac{{2\Delta \theta }}{{Z - 1}},z \in \left\{ {1,2,...,Z} \right\}$. The optimization problem can be formulated as
\begin{subequations}
\begin{align}\label{OP_imperfect1}\
&\mathop {\max }\limits_{\left\{ {{\bf{w}}\left( t \right), {\bf{x}}, {\bf{f}}  } \right\}} {\rm{ }}\left\{ {{R_b}\left( t \right) - \mathop {\max }\limits_{Y,Z} {R_{{e_{y,z}}}}\left( t \right)} \right\}\\
&{s.t.} \quad {\left\| {\bf{w}}\left( t \right) \right\|^2} \le {P_{max}},\label{power_constraintop_imperfect1}\\
&\quad \quad {x_1} = 0,{x_M} \le {D_{\max }}, \label{distance_constraint1op_imperfect1}\\
&\quad \quad {x_m} - {x_{m - 1}} \ge {D_0}, \forall m, \label{distance_constraint2op_imperfect1}\\
&\quad \quad {f_C} \le {f_m} \le {f_C} + \Delta F, \forall m .\label{frequency_constraintop_imperfect1}
\end{align}
\end{subequations}
Problem (30) is non-convex function with the coupled variables which is challenging to solve. Inspired by {\emph{Remark 2}}, we conclude that the optimization for ABV cannot affect that for APV and AFV. Thus, the proposed two-stage AO algorithm can effectively solve it.

\subsection{Optimizing ${\bf{w}}\left( t \right)$ with Given ${\bf{x}}$ and ${\bf{f}}$ }
 For given APV and AFV, problem (29) can be converted to
 \begin{subequations}
\begin{align}\label{OP_imperfect2}\
&\mathop {\max }\limits_{ \bf{w} } {\rm{ }}\left\{ {{{\log }_2}\left( {1 + {{\bf{w}}^H}{\bf{Aw}}} \right) - \mathop {\max }\limits_{Y,Z} {{\log }_2}\left( {1 + {{\bf{w}}^H}{{\bf{B}}_{y,z}}{\bf{w}}} \right)} \right\}\\
&{s.t.} \quad {\left\| {\bf{w}}\right\|^2} \le {P_{max}}.\label{power_constraintop_imperfect2}
\end{align}
\end{subequations}
We define a matrix as ${\bf{W}} = {\bf{w}}{{\bf{w}}^H}$ which follows that  ${\rm{rank}}\left( {\bf{W}} \right) = 1$, and the rank-1 constraint is non-convex which can be relaxed by the semi-definite relaxation (SDR) method. Then, problem (31) can be converted to
 \begin{subequations}
\begin{align}\label{OP_imperfect3}\
&\mathop {\max }\limits_{\bf{W}} \left\{ {{{\log }_2}\left( {1 + {\bf{Tr}}\left( {{\bf{A}}{\bf{W}}} \right)} \right) - \mathop {\max }\limits_{Y,Z} {{\log }_2}\left( {1 + {\bf{Tr}}\left( {{{\bf{B}}_{y,z}}{\bf{W}}} \right)} \right)} \right\}\\
&{s.t.} \quad {\bf{Tr}}\left( {\bf{W}} \right) \le {P_{\max }}.\label{power_constraintop_imperfect3}
\end{align}
\end{subequations}
Problem (32) is still non-convex and difficult to solve, we next resort to Lemma 2 via Fenchel conjugate arguments \cite{Convex}.

{\emph{\textbf{Lemma 2:}}} Considering the function $\varphi \left( u \right) =  - uv + \ln u + 1$ for any $v>0$.
 \begin{align}\label{FCA}\
\mathop {\max }\limits_{u > 0} \varphi \left( u \right) =  - \ln v
\end{align}
with the optimum value $u^{opt} = 1/v$. By utilizing Lemma 2, we have
 \begin{align}\label{Rereformulation}\
\ln 2{\log _2}\left( {1 + Tr\left( {{{\bf{B}}_{y,z}}{\bf{W}}} \right)} \right) &= \ln \left( {1 + Tr\left( {{{\bf{B}}_{y,z}}{\bf{W}}} \right)} \right)\nonumber\\
 &= {u_e}\left( {Tr\left( {{{\bf{B}}_{y,z}}{\bf{W}}} \right) + 1} \right)\nonumber\\
 &- \ln {u_e} - 1.
\end{align}
where ${u_e} = {\left( {Tr\left( {{{\bf{B}}_{y,z}}{\bf{W}}} \right) + 1} \right)^{ - 1}}$. It is worth mentioning that ${u_e}$ and $\bf{W}$ are coupled optimization variables, and we apply the AO method to solve it. Specifically, with fixed $\bf{W}$, the expression of the optimal $u_e^{opt}$ can be given in closed form as
 \begin{align}\label{opt_ue}\
u_e^{opt} = {\left( {Tr\left( {{{\bf{B}}_{y,z}}{\bf{W}}} \right) + 1} \right)^{ - 1}}.
\end{align}
Then, with given ${u_e}$, the optimization problem can be formulated as
 \begin{subequations}
\begin{align}\label{OP_imperfect4}\
&\mathop {\max }\limits_{{\bf{W}},g}\quad \ln \left( {1 + {\bf{Tr}}\left( {{\bf{AW}}} \right)} \right) - g\\
&{s.t.} \quad \quad {u_e}\left( {Tr\left( {{{\bf{B}}_{y,z}}W} \right) + 1} \right) - \ln {u_e} - 1 \le g,\forall y,z,\label{power_constraintop_imperfect4}\\
&\quad \quad \quad{\bf{Tr}}\left( {\bf{W}} \right) \le {P_{\max }}.\label{power_constraintop_imperfect4}
\end{align}
\end{subequations}
Problem (36) is a semi-definite program (SDP) problem which can be efficiently solved by convex solvers such as CVX \cite{2008CVX}. We observe that the optimal solution ${{\bf{W}}^{opt}}$ may not satisfy the rank-1 constraint. Thus, we apply the Gussian randomization method to recover ${{\bf{w}}^{opt}}$ approximately \cite{8811733}.

\subsection{Optimizing ${\bf{x}}$ with Given ${\bf{w}}\left( t \right)$ and ${\bf{f}}$. }
Different from \eqref{w_opt}, we cannot get the expression of the optimal ABV in closed form from problem (36). Hence, there is no simplified optimization object when optimizing APV and AFV. Inspired by {\emph{Remark 1}}, we conclude that the optimization problems for APV and AFV are actually to minimize the inner product between the channels of Bob and Eve. Therefore, the optimization problem can be formulated as \cite{10097703}
\begin{subequations}
\begin{align}\label{OP2_imperfect}\
&\mathop {\min }\limits_{{{{\bf{x}}}}} \mathop {\max }\limits_{Y,Z} {\left| {\left\langle {{{\bf{h}}_{ab}},{{\bf{h}}_{ae}}\left( {{r_y},{\theta _z}} \right)} \right\rangle } \right|^2}\\
&{s.t.} \quad {x_1} = 0,{x_M} \le {D_{\max }}, \label{distance_constraint1}\\
&\quad \quad {x_m} - {x_{m - 1}} \ge {D_0}, \forall m, \label{distance_constraint2}
\end{align}
\end{subequations}
where
 $\mathop {\max }\limits_{Y,Z} {\left| {\left\langle {{{\bf{h}}_{ab}},{{\bf{h}}_{ae}}\left( {{r_y},{\theta _z}} \right)} \right\rangle } \right|^2}$ aims to consider the worst sample in the suspicious region of Eve. According to \eqref{OPAO2}, ${\left| {\left\langle {{{\bf{h}}_{ab}},{{\bf{h}}_{ae}}\left( {{r_y},{\theta _z}} \right)} \right\rangle } \right|^2}$ is not monotonic for $\bf{x}$ which means that we cannot find the worst sample directly. Thus, we propose to utilize the BCD method to search the worst sample and solve problem (37).

Specifically, in the $s$-th iteration of BCD, we first optimize ${x_{m}}$ where $m = s{\kern 1pt} {\kern 1pt} {\kern 1pt} {\kern 1pt} {\rm{mod}}{\kern 1pt} {\kern 1pt} {\kern 1pt} M$, and the optimization problem is rewritten as follows:
\begin{subequations}
\begin{align}\label{OP2_imperfect1}\
&\mathop {\min }\limits_{{x_{m}}} \mathop {\max }\limits_{Y,Z} { {\sum\limits_{\scriptstyle{n=0}\hfill\atop
\scriptstyle{n \ne  m}\hfill}^{M-1} {{\cos \left[ {2\pi \left( {{{\tau _{m,y,z}}}{f_{m}} - {\tau _{n,y,z}}{f_{n}}} \right)} \right]}} }}\\
&{s.t.} \quad {x_1} = 0,{x_M} \le {D_{\max }}, \label{distance_constraint1_imperfect}\\
&\quad \quad {x_m} - {x_{m - 1}} \ge {D_0}, \forall m, \label{distance_constraint2imperfect}
\end{align}
\end{subequations}
where ${\tau _{m,y,z}} = {x_m}\frac{{\sin {\theta _z} - \sin {\theta _b}}}{c} + \frac{{{r_y} - {r_e}}}{c}$. Here, we utilize the linear search method to solve this problem, and the search range for $x_m$ is $\left[ {x_{m - 1}^s + {D_0},{D_1}} \right]$. Then, we update ${{\bf{x}}^{s - 1}}$ with $x_m^{opt}$ where $x_m^{opt}$ is the optimal solution, i.e., ${{\bf{x}}^s} = {{\bf{x}}^{s - 1}}$.

\subsection{Optimizing ${\bf{f}}$ with Given ${\bf{w}}\left( t \right)$ and ${\bf{x}}$. }
Similar to problem (37), the optimization problem in this case is formulated as
\begin{subequations}
\begin{align}\label{OP2_imperfect_f}\
&\mathop {\min }\limits_{{{{\bf{f}}}}} \mathop {\max }\limits_{Y,Z} {\left| {\left\langle {{{\bf{h}}_{ab}},{{\bf{h}}_{ae}}\left( {{r_y},{\theta _z}} \right)} \right\rangle } \right|^2}\\
&{s.t.} \quad {f_C} \le {f_m} \le {f_C} + \Delta F, \forall m. \label{distance_constraint2_f}
\end{align}
\end{subequations}
We propose to utilize the BCD and linear search method to solve this problem, and the search range for $f_m$ is $\left[ {{f_C},{f_C} + \Delta F} \right]$.

{\emph{Remark 3:}}
According to \eqref{habwithhae}, it is clear that ${{{\left| {\left\langle {{{\bf{h}}_{ab}},{{\bf{h}}_{ae}}\left( {{r_y},{\theta _z}} \right)} \right\rangle } \right|}^2}}$ is not monotonic for ${r_y}$ and ${\theta _z}$, which means that we cannot directly find the worst-case coordination for Eve in the estimated region $\left( {\cal{R}},{\cal{A}} \right)$. To solve it, we utilize the discretization method to simplify the region to some samples and search the worst case when optimizing APV and AFV. This method can effectively to deal with the problems of the position uncertainty in FDA \cite{10097703}. It is worth mentioning that multiple samples can make the results accurate, but it also leads to the complexity of the optimization problem.

\subsection{Overall Two-Stage AO Algorithm}

We propose to utilize the two-stage AO algorithm to solve the problem (30). Specifically, the APV and AFV at the transmitter are optimized in the first stage by utilizing the BCD method which is summarized in Algorithm 3. We consider the worst case among all cases and aim to minimize the inner product between the channels of Bob and Eve. In the second stage, we utilize the CVX method to obtain the optimal ABV which is summarized in Algorithm 4. The overall algorithm is the same as Algorithm 2.

\begin{algorithm}
\caption{Optimizing ${\bf{x}}$ by BCD}
\begin{algorithmic}[1]
\STATE {\bf{Initialization:}} Initializing ${\bf{f}}$ via Algorithm 3 and ${{\bf{w}}_{opt}}$ via Algorithm 4.
\STATE {\bf{Set:}} $s=0$.
\STATE {{\bf{Repeat:}}\\
        a) $s=s+1$, and $m = s{\kern 1pt} {\kern 1pt} {\kern 1pt} {\kern 1pt} {\rm{mod}}{\kern 1pt} {\kern 1pt} {\kern 1pt} {\rm{M}}$.\\
        b) Searching $x_m$ in the given region $\left[ {x_{m - 1}^s + {D_0},{D_1}} \right]$ to minimize $\mathop {\max }\limits_{Y,Z} {\left| {\left\langle {{{\bf{h}}_{ab}},{{\bf{h}}_{ae}}\left( {{r_y},{\theta _z}} \right)} \right\rangle } \right|^2}$.  \\
        c) We update ${{\bf{x}}^{s-1}}$ with $x_m^{opt}$ and obtain ${{\bf{x}}^s}$.\\}
\STATE {{\bf{Until}}  $\left| {{C_{\sec }}\left( {{{\bf{w}}_{opt}},{{\bf{x}}^s},{\bf{f}}} \right) - {C_{\sec }}\left( {{{\bf{w}}_{opt}},{{\bf{x}}^{s - 1}},{\bf{f}}} \right)} \right| \le \varsigma $ is satisfied.}
\end{algorithmic}
\end{algorithm}

The complexity of Algorithm 3 mainly depends on the number of iterations for the BCD method and samples for the coordinate uncertainty at Eve. We denote the increment of the position search as $\Delta x$, thus the complexity of the position search is ${\cal{O}}\left( {\frac{{{D_1} - x_{m - 1}^s - {D_0}}}{{\Delta x}}} \right)$. Moreover, $Y\times Z$ and ${\rm{I}}_{BCD}^{\bf{x}}$ denote the number of the samples and iterations required for BCD convergence, respectively. The complexity of Algorithm 3 is ${\cal{O}}\left( {\frac{{\left( {{D_1} - x_{m - 1}^s - {D_0}} \right)YZ}}{{\Delta x}}{{\rm{I}}_{BCD}^{\bf{x}}}} \right)$. When optimizing AFV, the complexity of Algorithm 3 is ${\cal{O}}\left( {\frac{{\Delta F}}{{\Delta f}}YZ{{\rm{I}}_{BCD}^{\bf{f}}}} \right)$.

\begin{algorithm}
\caption{Optimizing ${\bf{w}}$ by CVX and AO algorithm}
\begin{algorithmic}[1]
\STATE {\bf{Initialization:}} Initializing ${{\bf{w}}_0}$ according to the maximum transmit power constraint ${\left\| {{\bf{w}}_0}\right\|^2} \le {P_{max}}$.
\STATE {\bf{Set:}} $s=1$ and ${{\bf{W}}} = {\bf{w}}_0^H{{\bf{w}}_0}$.
\STATE {{\bf{Repeat:}}\\
        a) We obtain $u_{e,y,z}^{opt} = {\left( {Tr\left( {{{\bf{B}}_{y,z}}{\bf{W}}} \right) + 1} \right)^{ - 1}}$ and ${{\bf{u}}^s}= \left[ {u_{e,1,1}^{opt},...,u_{e,Y,Z}^{opt}} \right]$\\
        b) With given ${{\bf{u}}^s}$, we solve problem (36) by CVX and obtain the optimized ${\bf{W}}^s$. Then, we have $s = s + 1$.\\}
\STATE {{\bf{Until}}  $\left| {{C_{\sec }}\left( {{{\bf{W}}^s},{{\bf{u}}^s}} \right) - {C_{\sec }}\left( {{{\bf{W}}^{s - 1}},{{\bf{u}}^{s - 1}}} \right)} \right| \le \varsigma  $ is satisfied.}
\STATE {If rank$\left( {{{\bf{W}}^s}} \right)=1$, the optimal ABV is given by ${{\bf{w}}^{opt}} = {\rm{eig}}\left( {{{\bf{W}}^s}} \right)$ where eig$\left(  \cdot  \right)$ is eigenvector decomposition. If rank$\left( {{{\bf{W}}^s}} \right)\ne1$, we apply the Gussian randomization
method.}
\end{algorithmic}
\end{algorithm}

The complexity of Algorithm 4 is ${\cal{O}}\left( {{M^{3.5}}\log \left( {\frac{1}{\varsigma }} \right)I_{AO}^{\bf{W}}} \right)$ where ${I_{AO}^{\bf{W}}}$ denotes the number of the iterations required for AO convergence.

\section{Numerical Results}

In this section, numerical results for MFDA-aided MISO wireless communication are presented. Unless otherwise specified, the maximum transmission power is ${P_{max}} = 10$dBm and the carrier frequency is ${f_C} = 10$GHz. The noise variances at Bob and Willie are $\sigma^2_{b}=-100$dBm and $\sigma^2_{e}=-100$dBm, respectively. The position coordinates of Bob and Eve are $\left( {{r_b},{\theta _b}} \right) = \left( {1000{\rm{m}},{{30}^\circ }} \right)$ and $\left( {{r_e},{\theta _e}} \right) = \left( {1000{\rm{m}},{{35}^\circ }} \right)$. The minimum antenna distance is ${D_0} = \frac{\lambda }{2}$ where $\lambda  = \frac{c}{{{f_C}}}$ is the wavelength of MFDA. The movable antenna boundary is ${D_{\max }} = 30\lambda $ and the frequency increment range is $\Delta F = 1$ GHz. Moreover, the path loss factor is given as \cite{8811733}
\begin{align}\label{LFS}\
{\rm{Lfs}}\left( r \right) = C{\left( {\frac{r}{R}} \right)^{ - \alpha }},
\end{align}
where $C=-30$dBm is the path loss at the reference distance $R=1{\rm{m}}$ and $r$ is the individual link distance. $\alpha$ is the path loss exponent where ${\alpha _{ab}} = 2$ and ${\alpha _{ae}} = 3$. Besides, the whole time is $T=1s$ and CRF is $\Delta {T_{CRF}}={10^{ - 6}}s$.

For comparison, the security performance of three benchmark strategies are also investigated as follows: 1) PA, i.e., only optimizing the ABV at Alice. 2) FDA, i.e., optimizing the ABV and AFV at Alice. 3) Upper bound, i.e., ${C_{\sec }} = {\log}\left( {1 + \sigma _b^{ - 2}{\rm{Lf}}{{\rm{s}}^2}\left( {{r_b}} \right){P_{\max }}M} \right)$.

\subsection{MFDA with Perfect Eve CSI}

\begin{figure}[ht]
\centering
{\includegraphics[width=0.53\textwidth]{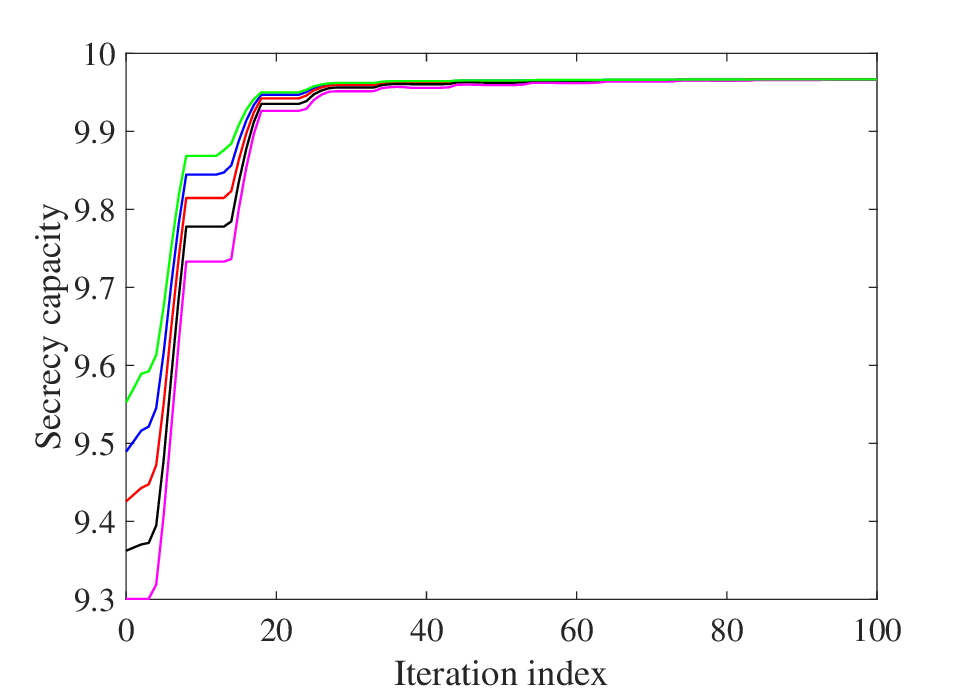}}
\caption{Typical converging traces of the BSUM for optimizing APV with $M=10$.}
\label{Fig2}
\end{figure}

Figure 2 depicts the convergence performance of the BSUM algorithm for APV where each trace starting from random initial APV. As expected, the secrecy capacity increases with the iteration index and converges at about 60 iterations which verifies the convergence of BSUM. Moreover, it is worth mentioning that the BUSM utilized for AFV has the same convergence curve as that for APV, thus we omit the figure here for brevity.

\begin{figure}[ht]
\centering
{\includegraphics[width=0.53\textwidth]{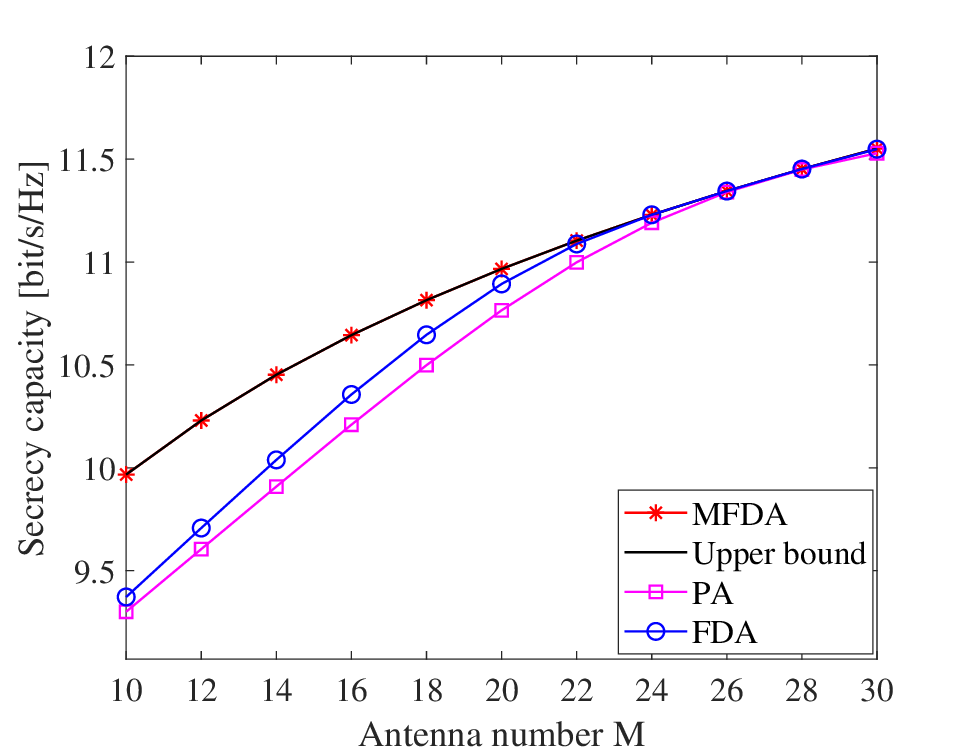}}
\caption{Secrecy capacity ${C_{\sec }}$ versus number of antennas $M$.}
\label{Fig3}
\end{figure}

Figure 3 depicts the secrecy capacity ${C_{\sec }}$ versus the number of antennas $M$ for different strategies. We observe that ${C_{\sec }}$ increases with $M$, because that brings more spatial DoFs. In particular, the secrecy capacity of MFDA is larger than that of PA and FDA, and can reach the upper bound with much fewer antennas. This is due to the fact that MFDA can provide larger secrecy capacity via adjusting the position and frequency of each antenna. However, FDA can only adjust the frequency, and is easily limited by the frequency constraint in a highly correlated channel scenario. This indicates that two-dimensional security achieved by MFDA can effectively handle different channel scenarios to enhance the security performance. Moreover, PA only optimizes the ABV at Alice which results in the worst security performance than other strategies as expected. Hence, PA requires more antennas to reach the upper bound.

\begin{figure}[ht]
\centering
{\includegraphics[width=0.53\textwidth]{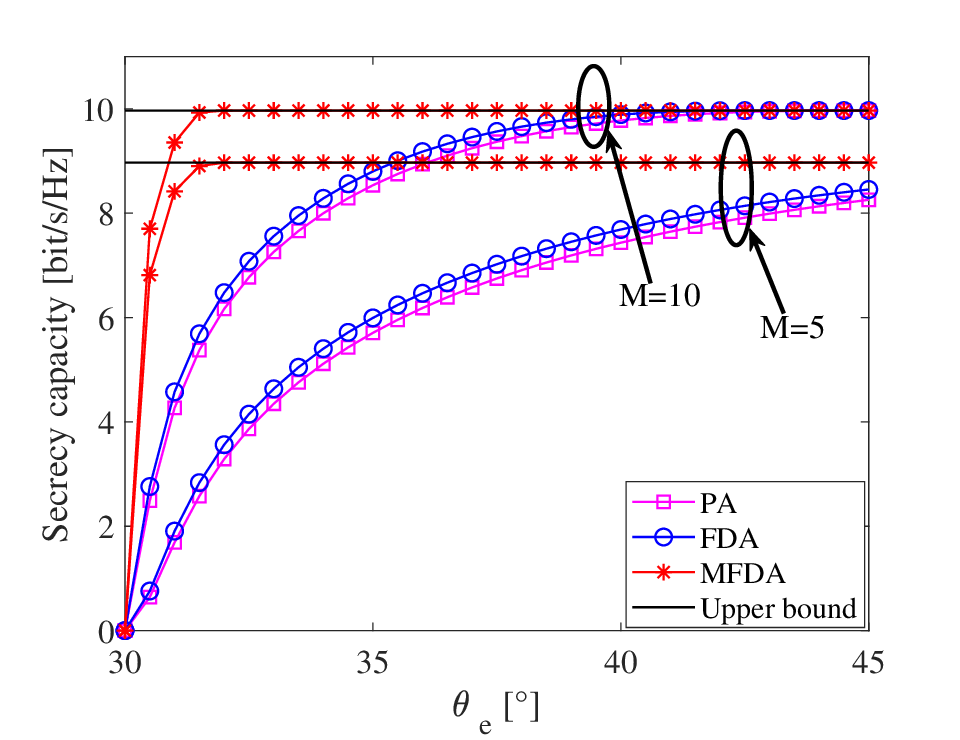}}
\caption{Secrecy capacity ${C_{\sec }}$ versus different Eve's angel ${\theta _e}$.}
\label{Fig4}
\end{figure}

Figure 4 illustrates how different angles ${\theta _e}$ affect the secrecy capacity ${C_{\sec }}$. We first observe that ${C_{\sec }}$ increases with ${\theta _e}$ and gradually approaches the upper bound.
It is because that, when Eve approaches Bob, the channels of Bob and Eve are highly correlated which increases the difficulty of channel decoupling, thus resulting in a corresponding decrement in secrecy capacity. In particular, when ${\theta _e} = {30^ \circ }$, neither MFDA, FDA nor PA can achieve any secrecy capacity. Fortunately, our proposed MFDA can fully utilize the frequency and position dimensions of antennas to reduce the performance loss, thus the secrecy capacity of MFDA drops sharply only when Eve and Bob are very close. This also implies that MFDA has stronger decoupling ability for highly correlated channel compared with FDA and PA.

\begin{figure}[ht]
\centering
{\includegraphics[width=0.53\textwidth]{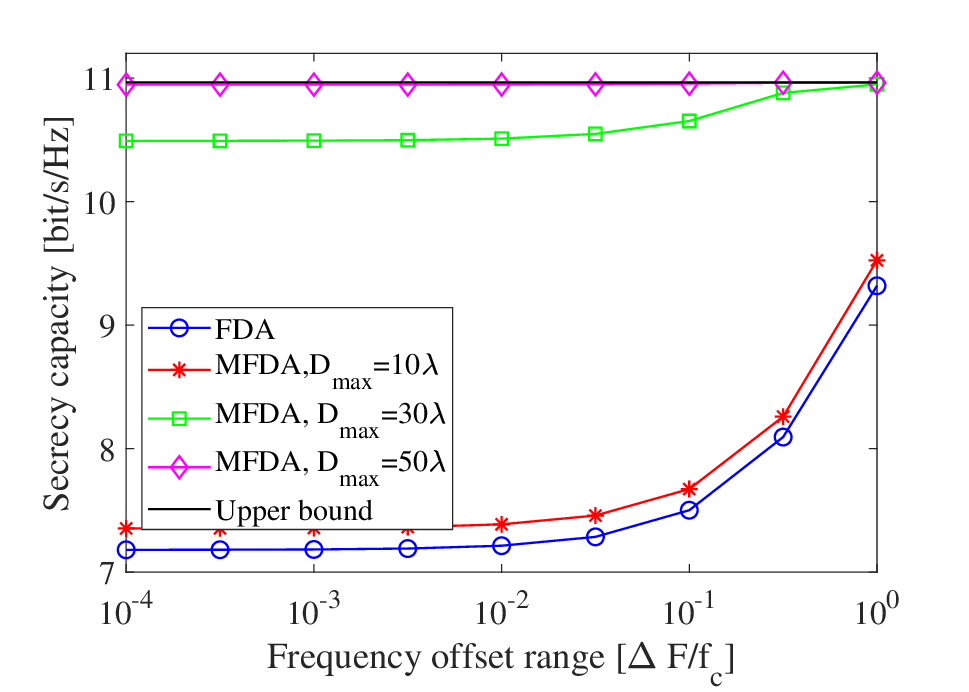}}
\caption{Secrecy capacity ${C_{\sec }}$ versus frequency increment range $\Delta F/{f_C}$ with $M=20$.}
\label{Fig5}
\end{figure}

Figure 5 analyzes the secrecy capacity ${C_{\sec }}$ for different frequency increment range $\Delta F/{f_C}$ and movable antenna boundary ${D_{\max }}$. It is clear that larger ${D_{\max }}$ and $\Delta F$ can provide more selectivity to optimize APV and AFV which can improve the secrecy capacity correspondingly. In practice, due to the device limitation or complex channel state, both APV and AFV will have constraints, such that relying on separately optimizing either of them cannot achieve the best security performance. Our proposed MFDA can jointly optimize APV and AFV to realize two-dimensional security which leads to the larger secrecy capacity than FDA and PA. In particular, when the AFV constraint is strict, MFDA can optimize APV to further improve secrecy performance instead of optimizing AFV.

\begin{figure}[ht]
\centering
{\includegraphics[width=0.53\textwidth]{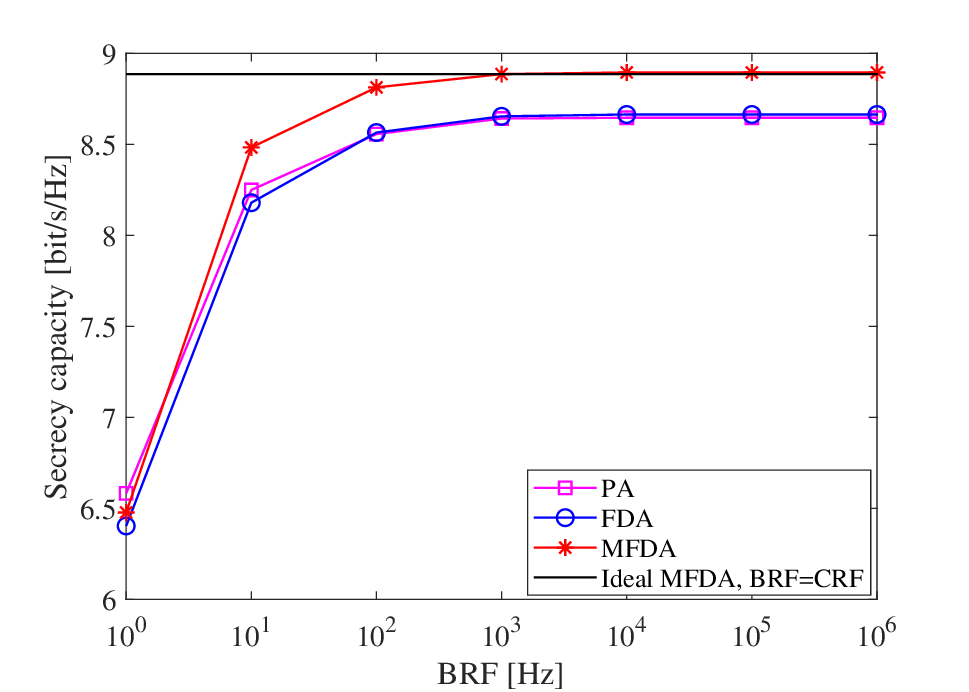}}
\caption{Secrecy capacity ${C_{\sec }}$ versus BRF with $M=20$, $CRF=1$MHz and $T=1$s.}
\label{Fig6}
\end{figure}

Figure 6 investigates the effects of BRF and CRF on secrecy capacity where the ideal MFDA means that the refresh speed of beamformer is larger than or equal to that of the channel. As expected, larger BRF can increase the secrecy capacity, and BRF $=1$KHz is sufficient to achieve the ideal upper bound. Moreover, when BRF $< {10^1}$Hz , PA outperforms FDA and MFDA. This is because a small BRF results in the current optimized ABV not being able to adapt to all the varying CSIs before the next optimization, thus cannot exploit the advantages provided by the frequency and position dimensions.

\subsection{MFDA with Imperfect Eve CSI}

In this subsection, we consider the case of imperfect CSI at Eve and investigate the influences of the uncertainties in angle and range on secrecy capacity.

\begin{figure}[htbp]
	\centering
	\begin{minipage}{0.49\linewidth}
		\centering
		\subfigure[BCD algorithm for APV]{\includegraphics[width=1.1\textwidth]{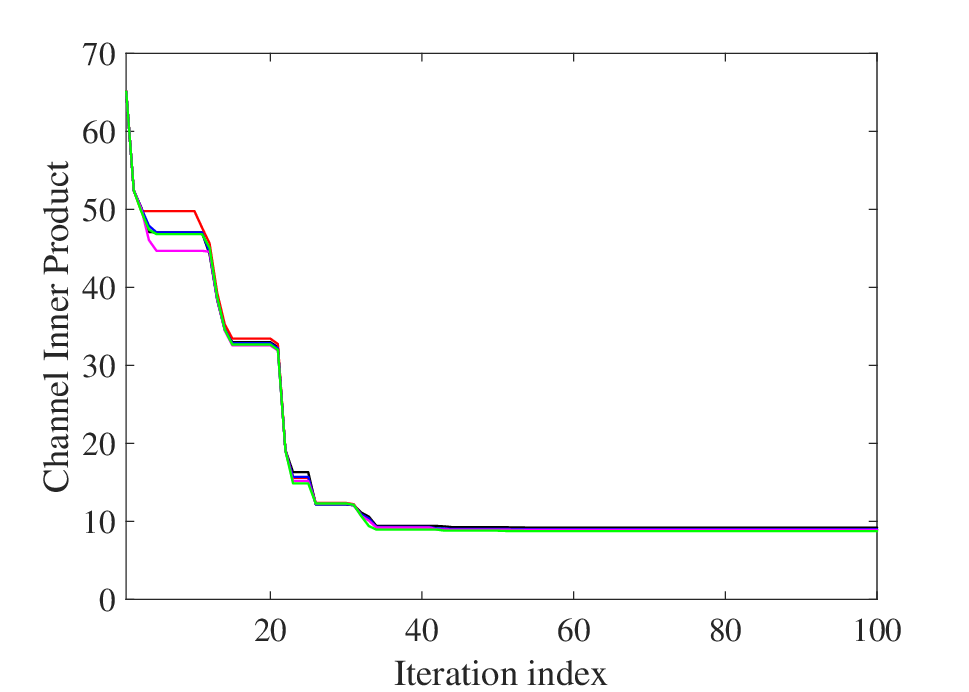}}
	\end{minipage}
	\begin{minipage}{0.49\linewidth}
		\centering
		\subfigure[AO algorithm for ABV]{\includegraphics[width=1.1\textwidth]{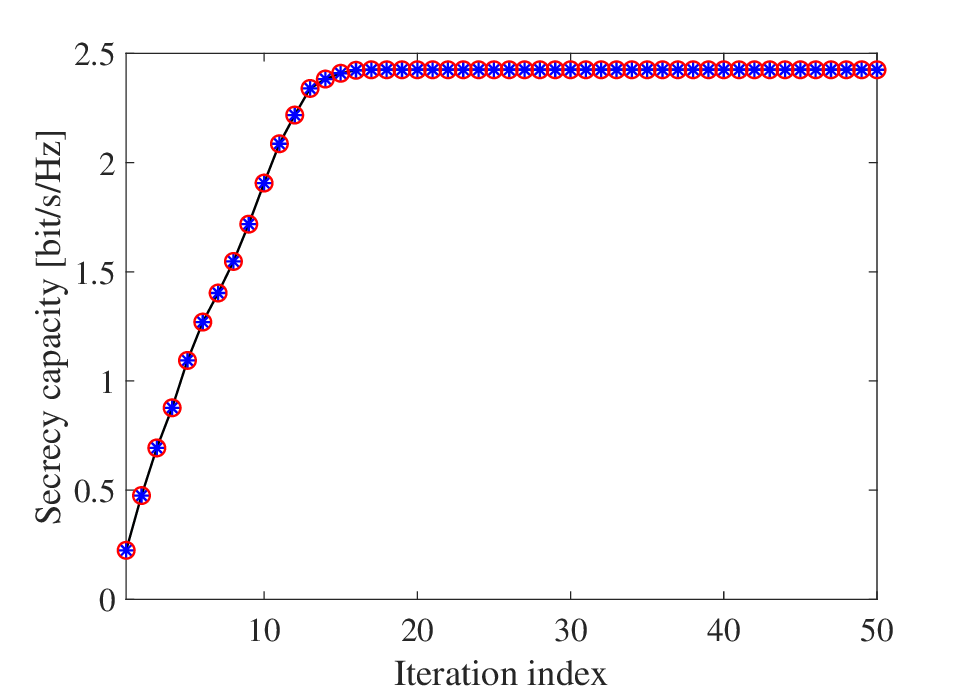}}
	\end{minipage}
\caption{Typical converging traces of the BCD algorithm and the AO algorithm based on CVX where $\Delta \theta  = {3^ \circ }$ and $Z=13$.}
\label{fig7}
\end{figure}

Figure 7(a) verifies the convergence of the BCD algorithm for optimizing APV where each trace starting from random initial APV and channel inner product is denoted as ${\left| {\left\langle {{{\bf{h}}_{ab}},{{\bf{h}}_{ae}}\left( {{r_y},{\theta _z}} \right)} \right\rangle } \right|^2}$. The converging behaviour for optimizing AFV is similar to that for optimizing APV. Thus, we omit the figure here for brevity. Figure 7(b) verifies the convergence of the AO algorithm based on CVX for optimizing ABV where each trace starting from random initial ABV subject to the maximum transmit power constraint.

\begin{figure}[htbp]
	\centering
	\begin{minipage}{0.49\linewidth}
		\centering
		\subfigure[Small uncertainty]{\includegraphics[width=1.1\textwidth]{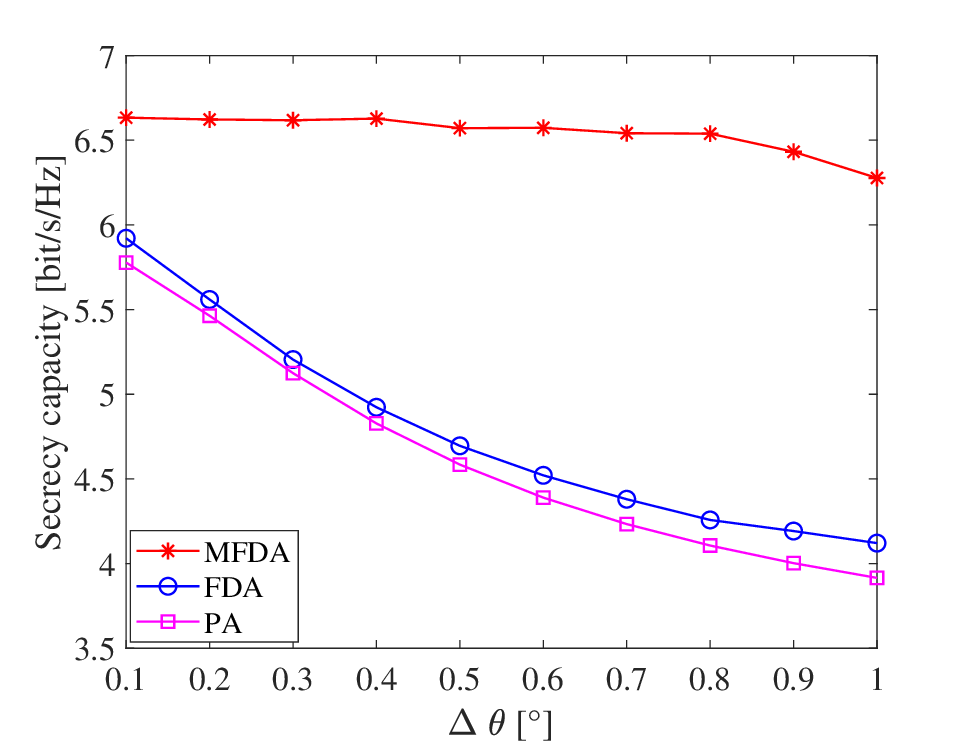}}
	\end{minipage}
	\begin{minipage}{0.49\linewidth}
		\centering
		\subfigure[Larger uncertainty]{\includegraphics[width=1.1\textwidth]{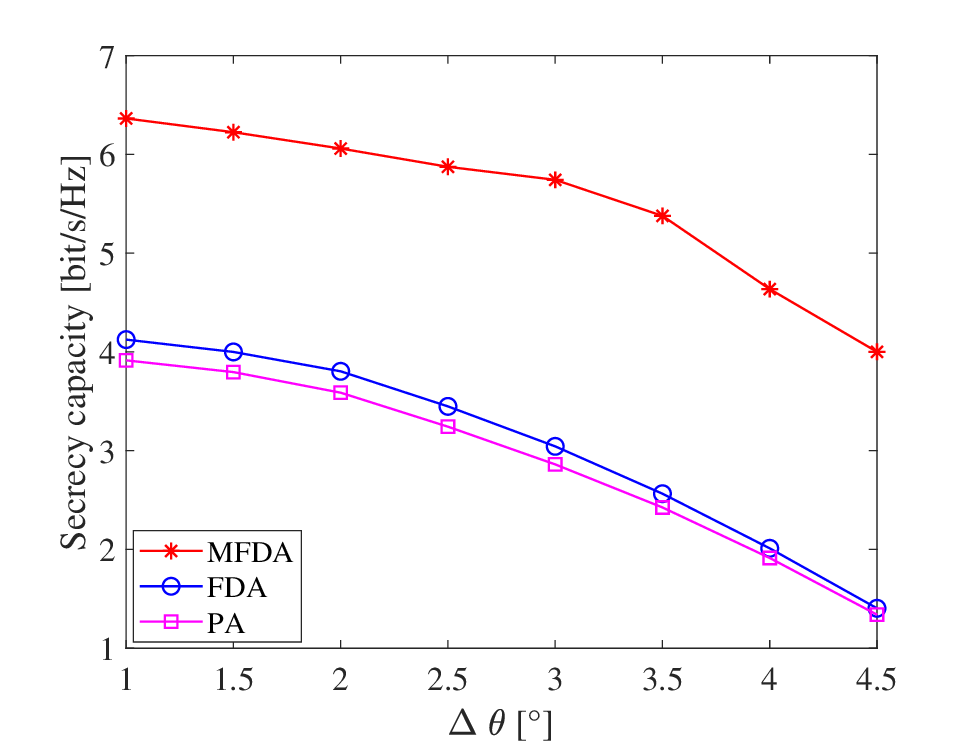}}
	\end{minipage}
\caption{Secrecy capacity ${C_{\sec }}$ versus angle uncertainty $\Delta \theta $ at Eve.}
\label{fig8}
\end{figure}

Figure 8 depicts the influences of angle uncertainty $\Delta \theta $ at Eve on secrecy capacity ${C_{\sec }}$. We consider two cases of angle uncertainty, i.e., small uncertainty and larger uncertainty. Specifically, small uncertainty implies that the estimated CSI at Eve has a small error and larger uncertainty implies that the estimated CSI has a larger error. It is clear that ${C_{\sec }}$ decreases with $\Delta \theta $ which means that the estimation error for Eve's CSI can deteriorate security performance, and larger estimation error leads to a more severe performance degradation. As expected, MFDA can significantly decrease performance loss caused by imperfect Eve CSI compared to FDA and PA. In particular, when the angle uncertainty is small, almost no loss in security performance by utilizing MFDA. Moreover, the influences of the range uncertainty $\Delta r$ at Eve on ${C_{\sec }}$ are the same as that of $\Delta \theta $, thus we omit the figures here for brevity.

\section{Conclusion}
In this paper, we proposed a novel antenna technology MFDA to enhance security performance in MISO wireless communication. Specifically, we aimed to maximize the secrecy capacity by jointly optimizing the ABV, APV and AFV at the transmitter, subject to the given constraints. Then, a two-stage AO algorithm with lower complexity was developed to solve the optimization problem. Moreover, we consider the case of imperfect CSI at Eve. Simulation results demonstrated that MFDA introduces a new dimension which can further improve security performance based on conventional FDA. Moreover, MFDA has strong channel identification ability to decouple the highly correlated channel, and can effectively reduce the security performance loss caused by CSI estimation error.

\begin{appendices}

\section{}

By substituting the optimal beamforming vector ${{\bf{w}}_{opt}}\left( t \right)$ into \eqref{OP1}, we have $\frac{{1 + {\bf{w}}_{opt}^H\left( t \right){\bf{A}}{{\bf{w}}_{opt}}\left( t \right)}}{{1 + {\bf{w}}_{opt}^H\left( t \right){\bf{B}}{{\bf{w}}_{opt}}\left( t \right)}} = {\lambda _{\max }}$ where ${\lambda _{\max }}$ is the maximum eigenvalue of ${\left( {{\bf{B}} + P_{\max }^{ - 1}{{\bf{I}}_M}} \right)^{ - 1}}\left( {{\bf{A}} + P_{\max }^{ - 1}{{\bf{I}}_M}} \right)$. By utilizing the matrix inverse lemma \cite{maxtrix}, we have
 \begin{align}\label{proof_1}\
&{\left( {{\bf{B}} + P_{\max }^{ - 1}{{\bf{I}}_M}} \right)^{ - 1}}\left( {{\bf{A}} + P_{\max }^{ - 1}{{\bf{I}}_M}} \right)\nonumber\\
 &= {P_{\max }}\left( {{{\bf{I}}_M} - \frac{{{P_{\max }}{\bf{h}}_{ae}^H{{\bf{h}}_{ae}}\sigma _e^{ - 2}}}{{\left( {1 + {P_{\max }}\left\| {{{\bf{h}}_{ae}}} \right\|_2^2} \right)}}} \right)\left( {\frac{{{\bf{h}}_{ab}^H{{\bf{h}}_{ab}}}}{{\sigma _b^2}} + {\frac{{{{\bf{I}}_M}}}{{{P_{\max }}}}} } \right)\nonumber\\
 &= {P_{\max }}\left( {\begin{array}{*{20}{l}}
{\frac{{{\bf{h}}_{ab}^H{{\bf{h}}_{ab}}}}{{\sigma _b^2}} - \frac{{{P_{\max }}{\bf{h}}_{ae}^H{{\bf{h}}_{ae}}{\bf{h}}_{ab}^H{{\bf{h}}_{ab}}}}{{\left( {1 + {P_{\max }}\left\| {{{\bf{h}}_{ae}}} \right\|_2^2} \right)\sigma _b^2\sigma _e^2}}}\\
{ - \frac{{{\bf{h}}_{ae}^H{{\bf{h}}_{ae}}}}{{\left( {1 + {P_{\max }}\left\| {{{\bf{h}}_{ae}}} \right\|_2^2} \right)\sigma _e^2}}}
\end{array}} \right) + {{\bf{I}}_M}.
\end{align}
Next, we decompose ${{\bf{h}}_{ab}}$ and ${{\bf{h}}_{ab}}$ into two orthogonal vectors
 \begin{align}\label{vector_decompose0}\
\frac{{{{\bf{h}}_{ab}}}}{{{\sigma _b}}} = \alpha {{\bf{h}}_e} + \beta {{\bf{h}}_{e \bot }},\frac{{{{\bf{h}}_{ae}}}}{{{\sigma _e}}} = {\left\| {{{\bf{h}}_{ae}}} \right\|_2}{{\bf{h}}_e} + 0 \cdot {{\bf{h}}_{e \bot }},
\end{align}
where ${{\bf{h}}_e}$ and ${{\bf{h}}_{e \bot }}$ are parallel and orthogonal unit vectors of ${{\bf{h}}_{ae}}$, respectively. In addition, $\alpha $ and $\beta $ are scalar values of ${{\bf{h}}_{ab}}$ subject to ${\alpha ^2} + {\beta ^2} = \frac{{\left\| {{{\bf{h}}_{ab}}} \right\|_2^2}}{{\sigma _b^2}}$ where ${\alpha ^2} = \frac{{{{\left| {\left\langle {{{\bf{h}}_{ab}},{{\bf{h}}_{ae}}} \right\rangle } \right|}^2}}}{{\sigma _b^2\left\| {{{\bf{h}}_{ae}}} \right\|_2^2}}$, ${\beta ^2} = \frac{{\left\| {{{\bf{h}}_{ab}}} \right\|_2^2}}{{\sigma _b^2}} - \frac{{{{\left| {\left\langle {{{\bf{h}}_{ab}},{{\bf{h}}_{ae}}} \right\rangle } \right|}^2}}}{{\sigma _b^2\left\| {{{\bf{h}}_{ae}}} \right\|_2^2}}$, $\left\| {{{\bf{h}}_{ab}}} \right\|_2^2 = {\rm{Lf}}{{\rm{s}}^2}\left( {{r_b}} \right)M$ and $\left\| {{{\bf{h}}_{ae}}} \right\|_2^2 = {\rm{Lf}}{{\rm{s}}^2}\left( {{r_e}} \right)M$.

By substituting \eqref{vector_decompose0} into \eqref{proof_1}, we have
 \begin{align}\label{proof_2}\
&\frac{{1 + {P_{\max }}\left\| {{{\bf{h}}_{ae}}} \right\|_2^2}}{{{P_{\max }}}}{\left( {B + P_{\max }^{ - 1}{{\bf{I}}_M}} \right)^{ - 1}}\left( {A + P_{\max }^{ - 1}{{\bf{I}}_M}} \right)\nonumber\\
 &= \left( {1 + {P_{\max }}\left\| {{{\bf{h}}_{ae}}} \right\|_2^2} \right)\frac{{{\bf{h}}_{ab}^H{{\bf{h}}_{ab}}}}{{\sigma _b^2}} - \frac{{{P_{\max }}{\bf{h}}_{ae}^H{{\bf{h}}_{ae}}{\bf{h}}_{ab}^H{{\bf{h}}_{ab}}}}{{\sigma _b^2\sigma _e^2}}\nonumber\\
 &- \frac{{{\bf{h}}_{ae}^H{{\bf{h}}_{ae}}}}{{\sigma _e^2}}\nonumber\\
& = \left( {{\alpha ^2} - \left\| {{{\bf{h}}_{ae}}} \right\|_2^2} \right){\bf{h}}_e^H{{\bf{h}}_e} + \alpha \beta {\bf{h}}_{e \bot }^H{{\bf{h}}_e} + \alpha \beta {\bf{h}}_e^H{{\bf{h}}_{e \bot }}\nonumber\\
 &+ {\beta ^2}{\bf{h}}_{e \bot }^H{{\bf{h}}_{e \bot }} + \left( {{P_{\max }^{ - 1}} + \left\| {{{\bf{h}}_{ae}}} \right\|_2^2} \right){{\bf{I}}_M}\nonumber\\
& = \left[ {{\bf{h}}_{e \bot }^H,{\bf{h}}_e^H} \right]{\bf{C}}\left[ {\begin{array}{*{20}{c}}
{{{\bf{h}}_{e \bot }}}\\
{{{\bf{h}}_e}}
\end{array}} \right],
\end{align}
where ${\bf{C}}$ is
 \begin{align}\label{proof_3}\
{\bf{C}} = \left( {P_{\max }^{ - 1} + \left\| {{{\bf{h}}_{ae}}} \right\|_2^2} \right){{\bf{I}}_2} + \left[ {\begin{array}{*{20}{c}}
{{\beta ^2}}&{\alpha \beta }\\
{\alpha \beta }&{{\alpha ^2} - \left\| {{{\bf{h}}_{ae}}} \right\|_2^2}
\end{array}} \right].
\end{align}

According to the properties of matrix determinant \cite{maxtrix}, we have $\det \left( {{\lambda _{\max }}{{\bf{I}}_2} - {\bf{C}}} \right) = 0$, i.e.,
 \begin{align}\label{proof_4}\
&\left( {{\lambda _{\max }} - \left( {P_{\max }^{ - 1} + \left\| {{{\bf{h}}_{ae}}} \right\|_2^2 + {\beta ^2}} \right)} \right)\nonumber\\
 &\times \left( {{\lambda _{\max }} - \left( {P_{\max }^{ - 1} + {\alpha ^2}} \right)} \right) - {\alpha ^2}{\beta ^2} = 0.
\end{align}
Then, the closed form expression of ${{\lambda _{\max }}}$ is given by
 \begin{align}\label{proof_5}\
{\lambda _{max}} & \buildrel \Delta \over =  \frac{{{P_{\max }}\left( {2P_{\max }^{ - 1} + \left\| {{{\bf{h}}_{ae}}} \right\|_2^2 + \left\| {{{\bf{h}}_{ab}}} \right\|_2^2 } \right)}}{{2\left( {1 + {P_{\max }}\left\| {{{\bf{h}}_{ae}}} \right\|_2^2} \right)}}\nonumber\\
 &+ \frac{{{P_{\max }}\sqrt {{{\left( {\left\| {{{\bf{h}}_{ab}}} \right\|_2^2 + \left\| {{{\bf{h}}_{ae}}} \right\|_2^2} \right)}^2} - 4\left\| {{{\bf{h}}_{ae}}} \right\|_2^2{\alpha ^2}} }}{{2\left( {1 + {P_{\max }}\left\| {{{\bf{h}}_{ae}}} \right\|_2^2} \right)}}.
\end{align}
It is clear that ${\lambda _{max}}$ decreases with ${{\alpha ^2}}$. Problem (11) aims to maximize ${\lambda _{max}}$ which is equivalent to minimizing ${{\alpha ^2}}$. Moreover, we note that ${\left| {\left\langle {{{\bf{h}}_{ab}},{{\bf{h}}_{ae}}} \right\rangle } \right|^2} = {\alpha ^2}\left\| {{{\bf{h}}_{ae}}} \right\|_2^2 = {\alpha ^2}M$ and ${\left| {\left\langle {{{\bf{h}}_{ab}},{{\bf{h}}_{ae}}} \right\rangle } \right|^2}$ monotonically increases with ${\alpha ^2}$, thus maximizing ${\lambda _{max}}$ is equivalent to minimizing ${\left| {\left\langle {{{\bf{h}}_{ab}},{{\bf{h}}_{ae}}} \right\rangle } \right|^2}$. Overall, problem (11) can be converted to problem (13).

\section{Proof of Lemma 1 }
\subsection{Case 1}

In this case, we have $y_{m,n}^{'}{\left( {\tau_m^{s - 1};\tau_n^{s - 1}} \right)} > 0$, ${y_{m,n}}\left( {{\zeta _{m,n}};\tau_n^{s - 1}} \right) =  - 1$ and ${\zeta _{m,n}} < \tau_m^{s - 1}$. The relation $-1 < {y_{m,n}}{\left( {\tau_m^{s - 1};\tau_n^{s - 1}} \right)} < 1$ holds if and only if there exists an integer ${\kappa _1} \ge 0$ such that
\begin{align}\label{Lemma1_case1}\
\begin{array}{l}
 - 1 < \cos \left[ {2\pi \left( {\tau_m^{s - 1}{f_m} - \tau_n^{s - 1}{f_n}} \right)} \right] < 1\\
\begin{array}{*{20}{l}}
\mathop  \Rightarrow \limits^{\left( a \right)} \pi  + 2\pi {\kappa _1} < 2\pi \left( {\tau_m^{s - 1}{f_m} - \tau_n^{s - 1}{f_n}} \right) < 2\pi  + 2\pi {\kappa _1}\\
\Rightarrow \frac{{2\left( {\tau_m^{s - 1}{f_m} - \tau_n^{s - 1}{f_n}} \right) - 2}}{2} < {\kappa _1} < \frac{{2\left( {\tau_m^{s - 1}{f_m} - \tau_n^{s - 1}{f_n}} \right) - 1}}{2}\\
{\mathop  \Rightarrow \limits^{\left( b \right)} {\kappa _1} = \frac{{\left\lfloor {2\left( {\tau_m^{s - 1}{f_m} - \tau_n^{s - 1}{f_n}} \right)} \right\rfloor  - 1}}{2}  },
\end{array}
\end{array}
\end{align}
where $\left( a \right)$ is due to ${\zeta _{m,n}} < \tau_m^{s - 1}$ and $\left( b \right)$ is because that ${\kappa _1}$ is an integer solution. Owing to ${y_{m,n}}\left( {{\zeta _{m,n}};\tau_n^{s - 1}} \right) =  - 1$, we have
\begin{align}\label{Lemma_xi1}\
2\pi \left( {{\zeta _{m,n}}{\tau _m} - \tau_n^{s - 1}{f_n}} \right) = \pi  + 2{\kappa _1}\pi
\end{align}
Thus, ${\zeta _{m,n}}$ is given by
\begin{align}\label{Lemma_xi_case1}\
{\zeta _{m,n}} = \frac{{\left\lfloor {2\left( {\tau_m^{s - 1}{f_m} - \tau_n^{s - 1}{f_n}} \right)} \right\rfloor }}{{2{f_m}}} + \frac{{\tau_n^{s - 1}{f_n}}}{{{f_m}}}.
\end{align}
According to \eqref{newobject_constraints}, we can obtain ${k_{m,n}}$ and ${\delta _{m,n}}$ through simple mathematical manipulations which are given by
\begin{align}\label{Lemma_case1_k}\
{{k_{m,n}} = \frac{{ - \pi {f_m}\sin \left[ {2\pi \left( {{\tau_m^{s - 1}}{f_m} - \tau_n^{s - 1}{f_n}} \right)} \right]}}{{{\tau_m^{s - 1}} - {\zeta _{m,n}}}}}
\end{align}
and
\begin{align}\label{Lemma_case1_delta}\
{{\delta _{m,n}} = \cos \left[ {2\pi \left( {{\tau_m^{s - 1}}{f_m} - \tau_n^{s - 1}{f_n}} \right)} \right] - {k_{m,n}}{{\left( {{\tau_m^{s - 1}} - {\zeta _{m,n}}} \right)}^2}}.
\end{align}

\subsection{Case 2}
In this case, we have $y_{m,n}^{'}{\left( {\tau_m^{s - 1};\tau_n^{s - 1}} \right)} > 0$, ${y_{m,n}}\left( {{\zeta _{m,n}};\tau_n^{s - 1}} \right) =  - 1$ and $ \tau_m^{s - 1} < {\zeta _{m,n}}$. The relation $-1 < {y_{m,n}}{\left( {\tau_m^{s - 1};\tau_n^{s - 1}} \right)} < 1$ holds if and only if there exists an integer ${\kappa _2} \ge 0$ such that

\begin{align}\label{case2}\
\begin{array}{l}
 - 1 < \cos \left[ {2\pi \left( {\tau_m^{s - 1}{f_m} - \tau_n^{s - 1}{f_n}} \right)} \right] < 1\\
\begin{array}{*{20}{l}}
{ \Rightarrow 2\pi {\kappa _2} < 2\pi \left( {\tau_m^{s - 1}{f_m} - \tau_n^{s - 1}{f_n}} \right) < \pi  + 2\pi {\kappa _2}}\\
 \Rightarrow {\kappa _2} = \frac{{\left\lceil {2\left( {\tau_m^{s - 1}{f_m} - \tau_n^{s - 1}{f_n}} \right)} \right\rceil  - 1}}{2}.
\end{array}
\end{array}
\end{align}
According to ${y_{m,n}}\left( {{\zeta _{m,n}};\tau_n^{s - 1}} \right) =  - 1$, we have
\begin{align}\label{Lemma_case2_xi1}\
2\pi \left( {{\zeta _{m,n}}{f_m} - \tau_n^{s - 1}{f_n}} \right) = \pi  + 2{\kappa _2}\pi.
\end{align}
Thus, ${\zeta _{m,n}}$ is given by
\begin{align}\label{Lemma_xi_case2}\
{\zeta _{m,n}} = \frac{{\left\lceil {2\left( {\tau_m^{s - 1}{f_m} - \tau_n^{s - 1}{f_n}} \right)} \right\rceil }}{{2{f_m}}} + \frac{{\tau_n^{s - 1}{f_n}}}{{{f _m}}}.
\end{align}
Similarly, based on \eqref{newobject_constraints}, we can obtain ${k_{m,n}}$ and ${\delta _{m,n}}$ which are the same as \eqref{Lemma_case1_k} and \eqref{Lemma_case1_delta}, respectively.

\subsection{Case 3}

In this case, we have $y_{m,n}^{'}{\left( {\tau_m^{s - 1};\tau_n^{s - 1}} \right)} =0$ and ${y_{m,n}}{\left( {\tau_m^{s - 1};\tau_n^{s - 1}} \right)} = -1$ which means that ${\zeta _{m,n}} = \tau_m^{s - 1}$ and ${\delta _{m,n}} =  - 1$. To find ${k_{m,n}}$, we first introduce a small parameter $\kappa _3$ whose value approaches zero, i.e., ${\kappa _3} \to 0$. Then, at the point ${\tau_m^{s - 1} + {\kappa _3}}$, we have $y_{m,n}^{'}\left( {\tau_m^{s - 1} + {\kappa _3};\tau_n^{s - 1}} \right) = u_{m,n}^{'}\left( {\tau_m^{s - 1} + {\kappa _3};\tau_n^{s - 1}} \right)$. Since ${\zeta _{m,n}} = \tau_m^{s - 1}$, we can obtain
\begin{align}\label{minimumpoint2}\
{k_{m,n}} &= \frac{{ - \pi {f_m}\sin \left[ {2\pi \left( {\tau_m^{s - 1}{f_m} - \tau_n^{s - 1}{f_n}} \right) + 2\pi {\kappa _3}{f_m}} \right]}}{{{\kappa _3}}}\nonumber\\
 &\mathop  = \limits^{\left( c \right)}  \frac{{\pi {f_m}\sin \left( {2\pi {\kappa _3}{f _m}} \right)}}{{{\kappa _3}}}.
\end{align}
where $\left( c \right)$ is due to the sum difference of product method \cite{IS}. It is worth noting that $\mathop {\lim }\limits_{{\kappa _3} \to 0} \frac{{\pi {f_m}\sin \left( {2\pi {\kappa _3}{f_m}} \right)}}{{{\kappa _3}}} = 2{\pi ^2}f_m^2$, thus ${k_{m,n}}=2{\pi ^2}f_m^2$.

\subsection{Case 4}
In this case, we have $y_{m,n}^{'}{\left( {\tau_m^{s - 1};\tau_n^{s - 1}} \right)} =0$ and ${y_{m,n}}\left( {\tau_m^{s - 1};\tau_n^{s - 1}} \right) = 1$ which means that ${\zeta _{m,n}} = \tau_m^{s - 1}$ and ${\delta _{m,n}} =  1$. Similar to case 3, we have $y_{m,n}^{'}\left( {\tau_m^{s - 1} - {\kappa _4};\tau_n^{s - 1}} \right) = u_{m,n}^{'}\left( {\tau_m^{s - 1} - {\kappa _4};\tau_n^{s - 1}} \right)$, such that we have
\begin{align}\label{minimumpoint2}\
{k_{m,n}} &= \frac{{\pi {f_m}\sin \left[ {2\pi \left( {\tau_m^{s - 1}{f_m} - \tau_n^{s - 1}{f_n}} \right) - 2\pi {\kappa _4}{f_m}} \right]}}{{{\kappa _4}}}\nonumber\\
 &=  - \frac{{\pi {f_m}\sin \left( {2\pi {\kappa _4}{f_m}} \right)}}{{{\kappa _4}}} =  - 2{\pi ^2}f_m^2.
\end{align}

\end{appendices}

\bibliography{FDA_MA}

\end{document}